\documentclass[a4paper,fleqn,usenatbib]{mnras}
\usepackage[T1]{fontenc}
\usepackage{ae,aecompl}
\usepackage{graphicx}	
\usepackage{amssymb}	
\defcitealias{K16}{I}
\defcitealias{K18ap}{II}

\title[Whole-grain heating temperatures]{Chemical significance of different temperature regimes for cosmic-ray-induced heating of whole interstellar grains}

\author[J. Kalv\=ans \& J. R. Kalnin]{
Juris Kalv\=ans\thanks{E-mail: juris.kalvans@venta.lv},
Juris Roberts Kalnin\\
Engineering Research Institute "Ventspils International Radio Astronomy Centre" of Ventspils University College,\\
In$\check{z}$enieru 101, Ventspils, LV-3601, Latvia\\
}
\date{Accepted 2018 Month X. Received 2018 Month Y; in original form 2018 December 7}
\pubyear{2018}

\begin{document}
\label{firstpage}
\pagerange{\pageref{firstpage}--\pageref{lastpage}}
\maketitle

\begin{abstract}
Cosmic-ray-induced whole-grain heating induces evaporation and other processes that affect the chemistry of interstellar clouds. With recent data on grain heating frequencies as an input for a modified rate-equation astrochemical model, this study examines, which whole-grain heating temperature regime is the most efficient at altering the chemical composition of gas and ices. Such a question arises because low-temperature heating, albeit less effective at inducing evaporation of adsorbed species, happens much more often than high-temperature grain heating. The model considers a delayed gravitational collapse of a Bonnor-Ebert sphere, followed by a quiescent cloud core stage. It was found that the whole-grain heating regimes can be divided in three classes, depending on their induced physico-chemical effects. Heating to low-temperature thresholds of 27 and 30\,K induce desorption of the most volatile of species -- N$_2$ and O$_2$ ices, and adsorbed atoms. The medium-temperature thresholds 40, 50, and 60\,K allow effective evaporation of CO and CH$_4$, delaying their accumulation in ices. We find that the 40\,K threshold is the most effective cosmic-ray induced whole-grain heating regime because its induced evaporation of CO promotes major abundance changes also for other compounds. An important role in grain cooling may be played by molecular nitrogen as the most volatile of the abundant species in the icy mantles. Whole-grain heating determines the sequence of accretion for different molecules on to grain surface, which plays a key role in the synthesis of complex organic molecules.

\end{abstract}
\begin{keywords}
astrochemistry -- molecular processes -- ISM: cosmic rays -- ISM: molecules
\end{keywords}

\section{Introduction} \label{intrd}

Cosmic-ray (CR) induced whole-grain heating (WGH) is the result of collisions between interstellar dust grains and CR ions that has an observable effect on the chemistry of interstellar clouds. When a CR particle -- typically a relativistic ion -- happens to hit a grain, it raises grain temperature by tens of kelvins \citep{Leger85}. The grain spends only a short time -- often a fraction of a second -- in the elevated temperature before cooling down to the ambient dust temperature. Yet, the temporal heating facilitates traversing of energy barriers for chemical reactions, diffusion and evaporation for molecules and atoms adsorbed on grain surfaces \citep{dHendecourt82,Hasegawa93}. Of these, evaporation has the greatest signature on interstellar cloud chemistry.

In nature, WGH occurs for interstellar grains that have been hit by CRs consisting of a variety of chemical element nuclei with different energies. CR particles usually pass through the grains, only those with energies below $\approx10$\,keV may be absorbed in the grains. Grains are being heated to different WGH temperatures $T_{\rm CR}$, and each $T_{\rm CR}$ has its own occurrence frequency $f_T$ \citep[hereinafter Paper~I]{K16}, which is equal to the rate at which CRs hit the grain, raising its temperature to $T_{\rm CR}$. WGH can be accurately described with the help of a temperature spectrum, where each $T_{\rm CR}$ has an attributed frequency $f_T$. Here we define $T_{\rm CR}$ at a point in time, when the energy from a CR hit has diffused within the whole volume of the grain, which thus has an uniform temperature and before any cooling processes have started to take effect. This time is about $10^{-9}$\,s after the CR hit \citep{Leger85}.

To the best of our knowledge, there has been no study or even an elaborate estimate comparing the chemical effects of different WGH temperatures. Nevertheless, as a physical process with consequences in chemistry, WGH has been included in many contemporary astrochemical studies \citep[see references in Section~1 of][hereinafter Paper II]{K18ap}. The primary $T_{\rm CR}$ regime used in astrochemical modelling is grain heating to 70\,K. This $T_{\rm CR}$ threshold was suggested by \citet{Hasegawa93}, probably because such high-temperature grains contain higher amount of thermal energy that can be used for rapid evaporation of volatiles, such as CO. Another $T_{\rm CR}$ threshold is 27\,K, the temperature required to induce chemical explosions in irradiated ices \citep{Greenberg73,dHendecourt82}. Grain heating to higher temperatures enables more efficient overcoming of energy barriers for molecules adsorbed on grain surface.

Recently, exhaustive studies calculating the WGH temperature spectrum in interstellar clouds have been published (Papers \citetalias{K16} and \citetalias{K18ap}), providing the frequencies $f_T$ for the whole range of $T_{\rm CR}$ possible for interstellar grains. These data show that grain heating to lower WGH temperatures happens more often than heating to higher temperatures. This is because, first, grains are heated more intensively by heavier elements, which are rarer. Second, grains best absorb energy from CRs that have energies in the range 0.2--10\,MeV\,amu$^{-1}$ (Paper~\citetalias{K16}). Considerable energy can also be input by CR ions with energies up to 200\,MeV\,amu$^{-1}$. The latter interval (10--200\,MeV\,amu$^{-1}$) has a greater total CR flux because it is larger and the flux density also is growing with CR energy in this region of CR spectra \citep{Padovani09,Chabot16}. In other words, the cross-correlation of CR energy loss function in grain material and CR energy spectra in clouds results in higher frequencies for grains heated to lower temperatures \citep[see also][]{deBarros11}. In addition, grain cooling at low temperatures happens slower than at high temperatures.

These aspects raise the question, whether the regime with the high $T_{\rm CR}$ of 70\,K \citep[as assumed by][]{Hasegawa93} really is the most efficient in transforming the chemistry of interstellar clouds. The data of Paper~\citetalias{K18ap} present an opportunity to compare the effects of different WGH temperature regimes with the help of astrochemical modelling. In this study, we aim to clarify the WGH temperature threshold(s) that most profoundly affect the chemistry of interstellar cloud cores. The tasks corresponding to this aim are: deriving analytical formulas for $f_T$ of different $T_{\rm CR}$ regimes from the WGH data from Paper~\citetalias{K18ap}; implementation of these data in an astrochemical model; calculations with the model; and comparison of modelling results to determine, which $T_{\rm CR}$ regime has the most significant effect on the abundances of key chemical species in gas and ice.

The main novelty of this study arises with the first application of the WGH frequencies from Paper~\citetalias{K18ap} in an astrochemical model. Clarifying the effects of different WGH regimes is an essential step for understanding the significance of CR-grain interactions. Summarizing the main results of Paper~\citetalias{K18ap} in analytical formulas will also make our data more accessible to modellers.

Naturally, the adoption of a single WGH temperature is an approximation for the purpose of astrochemical modelling. In fact, the grains are affected by all $T_{\rm CR}$ regimes; moreover, low-temperature WGH arises from less energetic CR hits and from the cooling of higher-temperature grains. Modelling of such cascading and simultaneous processes is a complex task to be performed in a future study. In order to prepare for this task, in the results section we attempted to identify the main physico-chemical processes driving the changes in cloud chemistry, induced by each WGH temperature regime.

\section{Methods} \label{mtd}

The astrochemical model employed in this study is the latest iteration of the code \textsc{Alchemic-Venta} \citep{K15apj1}. An updated approach of WGH frequencies and detailed chemical processes in ices on the heated grains were added to the code.

\subsection{Macrophysical model} \label{mdl-ph}

A suitable macrophysical setting is an important aspect of the model. Authors who have studied the chemical effects induced by CRs acting on interstellar grains \citep{Hasegawa93,Shen04,Reboussin14,K14} have used pseudo-time dependent models of a cloud with constant density. Similarly to those simulations, a quiescent stage of a dense cloud core was considered also here.

However, in this study, the quiescent stage was preceded by a first stage, where the modelled cloud core contracts. Such a stage was included to examine WGH effects in an environment with changing density. This is necessary because the intensity of CR particles and, thus, the rate of WGH events actually depend on the cloud column density towards the core. The contraction stage starts with hydrogen atom column density $N_{\rm H}$ and numerical density $n_{\rm H}$ characteristic to diffuse molecular gas. These values then increase to the values of the dense, quiescent stage.

\begin{figure}
 \hspace{-1.0cm}
	\includegraphics[width=13cm]{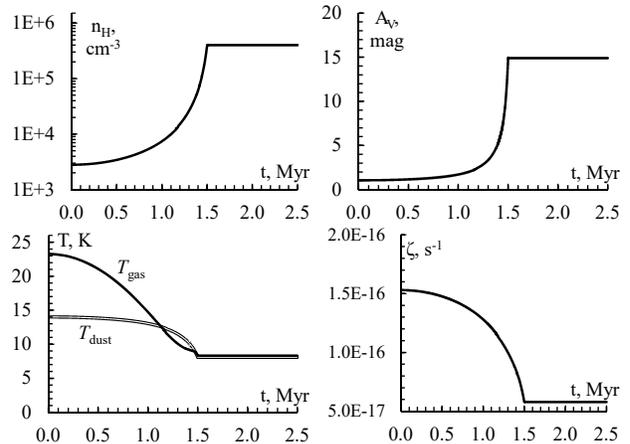}
 \vspace{-11.5cm}
    \caption{Physical conditions in the modelled central parcel of the cloud core.}
    \label{att-phys}
\end{figure}
The macrophysical model calculates $N_{\rm H}$ towards the centre part of a molecular cloud core. $N_{\rm H}$ was then used for calculating global physical parameters of the modelled gas parcel -- interstellar extinction $A_V$, temperature, and rates of CR-induced processes. The cloud core was assumed to be surrounded by diffuse gas with isotropic column density $N_{\rm out}=5\times10^{20}$\,cm$^{-2}$.

During the first stage, the model traces WGH-induced changes in the chemistry under evolving physical conditions. A 4\,M$_{\sun}$ Bonnor-Ebert sphere supposedly undergoes a delayed gravitational collapse. The collapse determines the central numerical density of all H atoms \citep{Nejad90}, starting with $n_{\rm H}=2800$\,cm$^{-3}$ and increasing to $4\times10^5$\,cm$^{-3}$ over an integration time $t=1.5$\,Myr. Other core parameters were calculated from $n_{\rm H}$ following Equations (1) and (2) of \citet{K18mn}.

The radius of the contracting sphere decreases from $5.1\times10^4$ to $3.0\times10^4$\,AU, while $N_{\rm H}$ toward the central parcel of gas increases from $2.1\times10^{21}$ to $3.0\times10^{22}$\,cm$^{-2}$. The initial conditions depict a large and rather diffuse core. Such a size means that its mass is spread over a large volume, resulting in a low initial column density towards the centre of the core, despite a considerable total mass and central density. These geometrical considerations have little effect on the present 0D chemical model, where only the value of $A_V$ is of importance.

The final value of $N_{\rm H}$ is higher than that typical for starless cores \citep[e.g.][]{Lippok13} and was chosen to enable analysing results for a medium, where the interstellar radiation field (ISRF) is attenuated and does not hamper the interpretation of WGH-induced chemical effects. In particular, this ensures that desorption is dominated by CR-induced effects.

In the second stage, which lasts another 1.0\,Myr, $n_{\rm H}$ and $N_{\rm H}$ are retained constant and equal to those at the end point of the first, contraction stage. This allows to trace chemistry under steady-state conditions. Figure~\ref{att-phys} shows the evolution of $n_{\rm H}$ and $A_V$, which was calculated as $N_{\rm H}/2\times10^{21}$.

The cloud collapse and quiescent core scenarios were intended for simulating conditions encountered in the interstellar medium, relevant to some extent for a variety of objects. The core was assumed to be part of a massive interstellar cloud, where most such cores have been observed. Such a core is partially shielded from incoming radiation by the proximity of sizeable interstellar gas reserves. More precisely, it was assumed that the ISRF photons (and CRs, Section~\ref{dffr}) can enter the cloud only from two directions, i.e., the core resides in a plane-parallel sheet, irradiated from both sides. This means that the ISRF intensity is reduced by a factor $g/4\pi$, where $g=2$ is the number of directions the cloud core is exposed to external irradiation. A factor $g=4\pi$ would characterize a lone, isolated spherical cloud.

During the collapse period, $A_V$ to the centre of the core changes from 1.05 to 14.9\,mag.  This increase means that the ISRF photons are replaced by CRs and CR-induced photons as the primary drivers of chemical processes in the cloud. The flux of the ISRF photons was taken to be $10^8$\,cm$^{-2}$\,s$^{-1}$, modified by the factor $g/4\pi$ because of geometric effects (see the paragraph above). This corresponds to $2.4\times10^{-4}$\,erg\,cm$^{-2}$\,s$^{-1}$. Shielding from the ISRF of H$_2$, CO and N$_2$ molecules was included with the help of tabulated data \citep{K15apj1}. Gas temperature $T_{\rm gas}$ \citep{K17} and dust temperature $T_{\rm dust}$ \citep{Hocuk17} were calculated as functions of $A_V$ (and thus, $N_{\rm H}$, see Figure~\ref{att-phys}). The model considers grains with radius $a=0.1$\,$\mu$m and density 3\,g\,cm$^{-3}$ that constitute 1 per cent of cloud mass.

\subsection{Chemical model} \label{mdl-ch}

\begin{table}
	\centering
\caption{Initial abundances of chemical species relative to H nuclei.}
\label{tab-elem}
	\begin{tabular}{lc}
  \hline
Species & Abundance \\
\hline
H$_2$ & 0.5 \\
He & 9.00E-2 \\
C & 1.40E-4 \\
N & 7.50E-5 \\
O & 3.20E-4 \\
F & 6.68E-9 \\
Na & 2.25E-9 \\
Mg & 1.09E-8 \\
Si & 9.74E-9 \\
P & 2.16E-10 \\
S & 9.14E-8 \\
Cl & 1.00E-9 \\
Fe & 2.74E-9 \\
\hline
	\end{tabular}
\end{table}
The model starts with the initial abundances of chemical species listed in Table~\ref{tab-elem}. With the given initial density, the calculated abundances of all chemical species converge to quasi-equilibrium values during the first 0.2\,Myr. This is well before an appreciable ice mantle with one monolayer (ML) thickness has formed on grain surfaces at $t=0.61$\,Myr. The latter is the first important reference point for a study focusing on surface chemistry.

For the gas chemical network, the model uses the UDfA12 database, which includes binary reactions, ISRF-induced photoreactions, CR-induced ionization and photoreactions \citep{McElroy13}. The surface chemical network is based on database for complex organic molecules (COMs) by \citet{Garrod08}. It is a reduced version because UDfA12 does not include so many organic species as the COMs network \citep[see also][]{K18mn}.

The UDfA12 does not discern between the isomeric radicals CH$_2$OH and CH$_3$O, while the COMs network does. Discerning between CH$_2$OH and CH$_3$O is important for the synthesis of methanol, a major ice molecule \citep[e.g.,][]{Garrod08,Vasyunin17}. We included both radicals in the network. For gas-phase reactions in this regard, data not provided by UDfA12 was adopted from the COMs network.

Neutral species are adsorbed on the inert surface of the grains, forming an icy mantle. The sticking coefficients were calculated according to \citet{Thi10} for the light species H and H$_2$, and taken to be unity for all heavier species. The surface density of adsorption sites was taken to be $1.5\times10^{15}$\,cm$^{-2}$, consistent with an olivine grain \citep{Hasegawa92}. The mantle was described as consisting of four layers -- the surface, $\approx1$~monolayer ($3.5\times10^{-8}$\,cm) thick, and three subsurface bulk-ice layers \citep{K15apj1}. As the MLs accumulate, the surface area of the grain increases accordingly. Total ice thickness can reach a maximum of up to $\approx100$\,MLs.

\textsc{Alchemic-Venta} has been derived from the ALCHEMIC code, a rate equation model, where surface reactions are treated with the modified rate equation method \citep{Semenov10}. Bulk-ice binary chemical reactions are permitted \citep{K15apj1}. Molecular dissociation reactions by ISRF and CR-induced photons in the surface and bulk-ice layers were adopted from the gas phase network (UDfA12). Following the result of \citet{K18mn}, the photodissociation rates for icy molecules were modified by a factor of 0.3. This is because the efficiency of the photoprocess in ices is lower thanks to neighbouring molecules that help stabilizing excited species. Moreover, nearby dissociation fragments have the possibility of immediate recombination.

In addition to WGH, several surface molecule desorption mechanisms were considered. These are evaporation at the ambient grain temperature $T_{\rm dust}$, reactive desorption, and photodesorption by ISRF and CR-induced photons. The latter includes also photodissociative desorption. The total (intact and dissociated molecules) photodesorption yield was assumed to be $10^{-3}$ for most species; for species with available experimental data the yields are listed in Table~3 of \citet{K18mn}.

\subsection{Features different from previous model} \label{dffr}

In addition to a detailed approach on WGH, several other changes were introduced to the code of \citet{K18mn}. We employed a 0D (point) version of the model, although the previous iteration of \textsc{Alchemic-Venta} is 1D. This means that only the central parcel of a cloud core was considered. This change was made to clearly trace changes in specific aspects of cloud chemistry that would not be obvious in a model considering a whole line of sight. Because the central parcel somewhat represents the overall composition of ices along the line of sight towards the centre of the core, a qualitative comparison with observations is possible.
cores.

Improvements were made to the micro-physical description of the icy mantles. Following \citet{Vasyunin17}, we assume that the ratio between surface binding energy $E_b$ and desorption energy $E_D$ is 0.55, and that the desorption energy of bulk-ice species is $2E_D$. The rate coefficient for diffusion of icy species between the four mantle layers was updated from that in the original \textsc{Alchemic-Venta} model of \citet[Equation~(11)]{K15apj1}. The diffusion rate coefficient is now
   \begin{equation}
   \label{dffr1}
k_{\rm diff} = \frac{R_{\rm hop}}{6l^2},
   \end{equation}
where $R_{\rm hop}$ is the molecule thermal hopping rate between two adjacent absorption in the icy mantle, $l$ is the average length of a molecule's track (in MLs) before diffusing into an adjacent ice layer and 6 is the assumed number of directions, where a molecule can diffuse to. This means that $k_{\rm diff}$ is now proportional to $l^{-2}$ (instead of $l^{-1}$), which is more physically correct. As in previous studies with this model, $l$ was taken to be half-thickness of the molecule's current ice layer.

CR-induced ionization rate and photon intensity were adjusted to ensure their compatibility with WGH rates in Section~\ref{htng}. CRs are attenuated by interstellar matter \citep{Indriolo09,Padovani09}. Taking this into account, the CR-induced ionization rate $\zeta$ (s$^{-1}$) was calculated as a function of $N_{\rm H}$ with initial spectra adopted from \citet{Ivlev15p}, model `High'. The same CR spectra was employed for the WGH rates in Papers \citetalias{K16} and \citetalias{K18ap}. To obtain $\zeta$, the calculated rate was modified by $g/4\pi$, much like the ISRF. Figure~\ref{att-phys} shows the resulting evolution of $\zeta$. The flux of CR-induced photons $F_{\rm CRph}$ is proportional to $\zeta$. Following \citet{Cecchi92}, it was taken to be 4875\,cm$^{-2}$\,s$^{-1}$ when $\zeta=1.7\times10^{-17}$\,s$^{-1}$. This means that
   \begin{equation}
   \label{dffr2}
F_{\rm CRph} = 2.868\times10^{20}\times\zeta\,.
   \end{equation}

\subsection{Whole-grain heating and associated processes} \label{supra}

\subsubsection{General considerations} \label{gnrl}

\citet{Hasegawa93} divided the time grains spend in the elevated WGH temperature with the time interval between energetic CR hits and found that grains spend a part of about $3.2\times10^{-19}$ of their lifetime in a heated state. With the conditions given in this paper, a 0.1\,M$_{\sun}$ interstellar gas parcel of a cloud contains about $1.6\times10^{44}$ 0.1$\mu$m grains. This means that about $5\times10^{25}$ grains would be in a heated state at any given instant. Compared to newer data, \citeauthor{Hasegawa93} underestimate WGH frequency for interstellar clouds. Also, grain heating to lower temperatures happens more often and the grains cool slower. For example, grain heating to 27\,K happens with a characteristic frequency $f_{>27}\approx10^{-10}$\,s$^{-1}$ (Section~\ref{htng}), while the cooling time-scale is about an hour (Section~\ref{clng}). This means that $6\times10^{37}$ grains have temperatures close to 27\,K at a given time instant. The WGH-affected grains cool, while others are heated, forming a continuously existing population of grains (and their icy mantles) that have a specific WGH temperature $T_{\rm CR}$.

The above means that, for the purpose of astrochemical modelling, the WGH-affected icy species can be considered as a separate ice phase. The whole WGH process is divided in three stages -- heating, molecular-level processes in the heated ices and cooling. The model borrows this approach from \citet{K15aa}. Molecules in each of the four ice layers at ambient temperature $T_{\rm dust}$ are converted to the same ice layer at WGH temperature $T_{\rm CR}$ with a first-order rate coefficient $k_{\rm heat}$. The icy molecules at $T_{\rm CR}$ are converted back to molecules in their respective ice layers with a cooling rate coefficient $k_{\rm cool}$. The abundance of species in the WGH phase is typically eight or more orders of magnitude lower than the abundance of the same species in its respective layer at ambient grain temperature. The coefficients $k_{\rm heat}$ and $k_{\rm cool}$ are discussed below in Sections \ref{ppri} and \ref{clng}. The molecules in the `heated ice' phase can diffuse between layers, evaporate and interact via chemical reactions, just like the icy species at the ambient temperature, with the only difference that their temperature is $T_{\rm CR}$ (Section~\ref{hotch}).

This approach on WGH heating is a significant improvement over the original approach of \citet{Hasegawa93} that simply included evaporation with an adjusted rate coefficient. For example, the rate of a process that occurs rapidly at the elevated temperature $T_{CR}$ is controlled by the grain heating and cooling rates, not the inherent rate of such a process.

To investigate the effectiveness of different WGH regimes, several WGH temperature thresholds were considered. These start with the 27\,K, the lowest $T_{\rm CR}$ threshold expected to induce significant changes in grain surface processes, and continue with 30, 40, 50, 60, 70, 80, and 90\,K thresholds. Together with `case zero' -- a model that does not include WGH -- this accounts for a total of nine simulations. As discussed in the results section~\ref{rslt}, the WGH thresholds can be divided in three classes, with a 2-3 simulations falling in each class. Therefore, the consideration of eight WGH regimes allows identification of all or most of the chemical mechanisms and their consequences, induced by WGH.

\subsubsection{Data of Paper\,\citetalias{K18ap}} \label{ppri}

The rate coefficient for ice molecule transition to the icy WGH phase is equal to the frequency of a single grain being heated to $T_{\rm CR}$, i.e., $k_{\rm heat}=f_T$. The new, comprehensive data of Paper~\citetalias{K18ap} allows to distinguish WGH regimes with different minimum $T_{\rm CR}$ thresholds. That study provides tabulated $f_T$ values for the full WGH temperature range for three grain sizes at conditions characteristic to starless cloud cores. These data were adapted for use in this study as described in Section~\ref{htng}.

For comprehensiveness and because the use of these data provides the main novelty in this paper, we briefly outline the principles used in Paper~\citetalias{K18ap} to obtain the WGH temperature spectrum. \citet{Hasegawa93} based their estimate of $f_{70}$ -- WGH frequency for $T_{\rm CR}\approx70$\,K -- on the frequency of iron CR collisions with interstellar grains. They $f_{70}$ derived from the data of \citet{Leger85}, who employed CR spectra derived in 1970s. More recent and precise interstellar CR energy spectra have been provided, e.g., by \citet{Moskalenko02,Indriolo09,Padovani09,Ivlev15p,Morlino15} and \citet{Chabot16}. Their results show that CR intensity drops with increasing column density of interstellar gas and that there is a relatively high flux of low energy CRs with energies $<100$\,MeV\,amu$^{-1}$. This had made necessary recalculating the WGH rates. Results of such studies \citep[Papers \citetalias{K16} and \citetalias{K18ap}]{Chabot16} showed that the grain heating rates can be vastly different from those suggested by \citet[see, e.g., the graphic comparison in Figure~9 of Paper~\citetalias{K18ap}]{Hasegawa93}.

Paper~\citetalias{K18ap} considers dust grains in a molecular cloud, irradiated by CRs with energies in the range 1\,eV...10\,GeV. The elemental composition of CRs, consisting of 32 different nuclei, was largely adopted from \textit{Voyager~1} data \citep{Cummings16}. The initial CR energy spectra \citep[from][]{Ivlev15p} was then modified for each nucleus according to the column density of gas traversed by the fast CR ions propagating through the molecular cloud core. In total, eight column densities of H atoms were considered with values relevant to molecular clouds and dense cores. Such dense regions are shielded from the ISRF, which allows accumulating a considerable ice layer on their grains. This changes grain heat capacity, i.e., CR hits with the same energy result in a $T_{\rm CR}$ that is lower for icy grains when compared to naked grain nuclei. This is an additional factor reducing the frequency of high-temperature WGH in dark cores. 

The result of the calculations in Paper~\citetalias{K18ap} is the frequency of CR hits that raise grain temperature to a certain temperature. To obtain the frequency $f_T$ for a minimum $T_{\rm CR}$ threshold, the frequencies for all temperatures above $T_{\rm CR}$ were summed up. For example, $f_{>70}$ is the frequency (s$^{-1}$) of events for a single grain having temperature higher than 70.00\,K. The summed frequencies for 0.1\,$\mu$m grains are listed in Table~4 of Paper~\citetalias{K18ap} and were used to derive handy analytical functions for the eight $T_{\rm CR}$ thresholds considered in this study (below).

\subsubsection{Rate of whole-grain heating by CRs} \label{htng}

\begin{figure}
 \vspace{-1.0cm}
 \hspace{-1.0cm}
	\includegraphics[width=11cm]{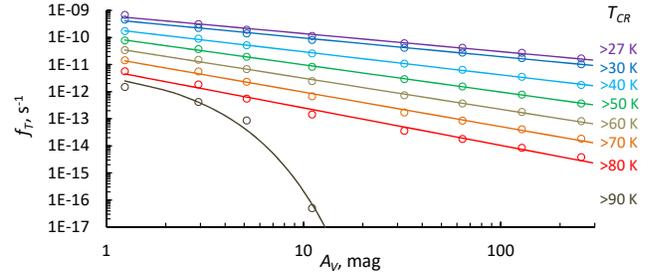}
 \vspace{-11cm}
    \caption{The frequencies $f_T$ for WGH events raising the temperature of a single interstellar dust grain above a certain threshold ($T_{\rm CR}$ regimes). Data points are from Paper~\citetalias{K18ap}, curves represent the analytical functions defined by Equations \ref{wgh1}--\ref{wgh7}.}
    \label{att-ft}
\end{figure}
We present analytical formulas from the data of Paper~\citetalias{K18ap} for the eight WGH temperature regimes. The frequencies $f_T$ are derived for 0.1\,$\mu$m grains shielded by column densities $2.5\times10^{21}...5.1\times10^{23}$\,cm$^{-2}$, corresponding to interstellar extinctions $A_V \approx 1.2...256$\,mag. The latter values are the maximum application limits for these data because at higher column densities the CR spectra are increasingly modified by spallation and other processes.

According to Paper~\citetalias{K18ap}, the $f_T$ functions are to be multiplied by the factor $g=2$, which indicates the extent of external shielding of CRs (Section~\ref{mdl-ph}). The calculated $f_T$ values are a function of hydrogen column densities, which are here expressed in terms of $A_V$. It was assumed that $N_{\rm H}/A_V=2.0\times10^{21}$\,cm$^{-2}$.

Figure~\ref{att-ft} pictures the $f_T$ functions for different WGH temperature regimes along with data points from Paper~\citetalias{K18ap}. $T_{\rm CR}$ in Paper~\citetalias{K18ap} has a built-in accuracy of 0.01; for example, $f_{>27}$ refers to $T_{\rm CR}>27.00$\,K. The analytical functions themselves are listed below -- frequency of WGH events for a single grain that lift its temperature above the $T_{\rm CR}$ threshold.
\begin{equation}
    T_{\rm CR}>27\,{\rm K};\,\, f_{>27}=g\times3.233\times10^{-10}A_V^{-0.6669}
	\label{wgh1}
\end{equation}
\begin{equation}
    T_{\rm CR}>30\,{\rm K};\,\, f_{>30}=g\times2.364\times10^{-10}A_V^{-0.6925}
	\label{wgh2}
\end{equation}
\begin{equation}
    T_{\rm CR}>40\,{\rm K};\,\, f_{>40}=g\times1.056\times10^{-10}A_V^{-0.8527}
	\label{wgh3}
\end{equation}
\begin{equation}
    T_{\rm CR}>50\,{\rm K};\,\, f_{>50}=g\times4.990\times10^{-11}A_V^{-1.007}
	\label{wgh4}
\end{equation}
\begin{equation}
    T_{\rm CR}>60\,{\rm K};\,\, f_{>60}=g\times2.207\times10^{-11}A_V^{-1.148}
	\label{wgh5}
\end{equation}
\begin{equation}
    T_{\rm CR}>70\,{\rm K};\,\, f_{>70}=g\times9.001\times10^{-12}A_V^{-1.273}
	\label{wgh6}
\end{equation}
\begin{equation}
    T_{\rm CR}>80\,{\rm K};\,\, f_{>80}=g\times3.036\times10^{-12}A_V^{-1.383}
	\label{wgh7}
\end{equation}
\begin{equation}
    T_{\rm CR}>90\,{\rm K};\,\, f_{>90}=g\times4.670\times10^{-12}{\rm exp}(-1.069A_V)
	\label{wgh8}
\end{equation}

\subsubsection{Chemical processes in the heated ice phase} \label{hotch}

The effects of processes in CR-heated ices have been analysed before, for the 70\,K $T_{\rm CR}$ regime. \citet{Hasegawa93,Roberts07} and \citet{Iqbal18} have studied the effects of WGH-induced evaporation. The significance of diffusion that facilitates chemical reactions has been studied by \citet{Reboussin14}, while diffusion between ice surface and bulk phases has been considered by \citet{K14}. \citet{K15aa} analysed the effects of diffusion and reactions, together with CR-induced dissociation of icy species. These studies show that processes other than evaporation affect the abundances of major species only by a few per cent. COMs and other organic molecules can be affected more significantly because of their low absolute abundances.

The approach on the modelling of WGH-induced chemistry was adapted from \citet{K15aa} and attributed to the WGH regimes with different $T_{\rm CR}$. As a separate phase, the species constituting WGH-affected ices have their own physical and chemical processes included in the model. These are evaporation, diffusion between the ice layers, and binary reactions. The rates of these processes were calculated in the same manner as for the icy species, except that they occur at the temperature $T_{\rm CR}$. For the binary reactions, the same chemical network, based on the COMs database, was used (Section~\ref{mdl-ch}). The reactions proceed via two steps -- the diffusion of reactants and overcoming the reaction activation energy barrier once the reactants have met. Both steps are affected by the elevated temperature. For surface reactions, reactive desorption was considered, too. Photochemistry of the icy molecules in the WGH phase was not included.

\subsubsection{Grain cooling} \label{clng}

A simple but comprehensive description for grain cooling was employed, suitable for calculating the duration of WGH events for an arbitrary temperature within the relevant $T_{\rm CR}$ interval (up to 90\,K). In the model, cooling is a first-order process that converts the molecules in the WGH ice phase with temperature $T_{\rm CR}$ back to the ice phase in the ambient grain temperature $T_{\rm dust}$. Cooling of the heated grains can occur via evaporation or electromagnetic radiation. Following \citet{Hasegawa93}, we assumed that evaporative cooling occurs with the time-scale of CO evaporation $t_{\rm evap}$. This includes an assumption that there is always a sufficient number of CO molecules on the surface to rapidly cool the grain. The adsorption energy of CO was taken to be 1150\,K, from the COMs chemical network. A more precise approach of evaporative grain cooling will be the subject of a dedicated future study.

Radiative grain cooling time-scales depend on grain size and other properties \citep{Duley73,Greenberg74,Aannestad79,Tabak87,Draine01,Cuppen06}. The radiative cooling time was estimated by integrating equation~(15) from \citet{Duley73} for a temperature interval spanning $0.1T_{\rm CR}$, centred on $T_{\rm CR}$. In other words, we assume that the grain resides at the temperature $T_{\rm CR}$ for the time that takes the grain to cool from temperature $1.05T_{\rm CR}$ to $0.95T_{\rm CR}$. Such an interval was chosen so that CO and N$_2$ evaporation rates (the most important processes, induced by WGH) do not deviate by more than about one order of magnitude from these rates at 27 or 30\,K at both ends of the interval. These low-$T_{\rm CR}$ regimes are those primarily affected by radiative cooling -- the time scale of radiative cooling becomes shorter than that of CO evaporation at $T_{\rm CR}\approx31$\,K.

Therefore, we obtain that
   \begin{equation}
   \label{cool1}
t_{\rm rad} = \frac{3Nk_B}{2a\sigma A} \times \left( \frac{1}{(0.95T_{\rm CR})^3}-\frac{1}{(1.05T_{\rm CR})^3} \right),
   \end{equation}
where $N$ is the number of particles in the grain, $k_B$ is the Boltzmann constant, $a$ is grain radius, $\sigma$ is the Stefan-Boltzmann constant, and $A$ is grain surface area. $N$ is estimated as the sum of the number of atoms in the amorphous grain nucleus and the number of molecules in the icy mantle. The inclusion of temperatures slightly higher than $T_{\rm CR}$ is adequate because $T_{\rm CR}$ is a minimum temperature threshold for WGH. The main actual effect of radiative cooling is that the energy grain receives from the impacting CR particle is not fully used for breaking bonds between evaporating molecules and grain surface. The total rate coefficient for the cooling process is
   \begin{equation}
   \label{cool2}
k_{\rm cool} = \frac{1}{t_{\rm evap}} + \frac{1}{t_{\rm rad}}.
   \end{equation}

\section{Results} \label{rslt}

To determine which $T_{\rm CR}$ regime has the greatest effect on cloud chemistry, the abundances of major species calculated with different $T_{\rm CR}$ regimes were compared to results of `model zero' that does not include WGH. The criterion for `major species' was that their abundance, relative to total hydrogen, ${\rm[X]}=n_{\rm X}/n_{\rm H}$ rises above $10^{-5}$ for at least one integration step. This criterion was met by CO, CO$_2$, H$_2$O, O$_2$, O, N$_2$, NH$_3$, and N in gas or ice phases. We divide these in oxygen species (Section~\ref{res-o}) and nitrogen species (\ref{res-n}). WGH affects also the abundances of more complex minor, organic species, either directly or by altering the proportions of their major parent molecules. This interesting aspect was covered in Sections \ref{res-cc} and \ref{res-com}.

The abundance curves in the results figures are shown starting with an integration time of 1.0\,Myr ($A_V=3.4$\,mag), when surface species have become abundant and WGH is able to influence a significant portion of heavy molecules. During early times, desorption is dominated by ISRF photons. Tables in Appendix~\ref{app-tab} present quantitative data -- the calculated abundances of the major species compared to the respective values from the `model zero'.

The WGH temperatures, represented in this study by the eight thresholds of $T_{\rm CR}$ (Section~\ref{gnrl}), can be divided in several classes. The low-temperature class includes the $>27$ and $>30$\,K regimes. It is efficient at evaporating the most volatile of all species, such as N$_2$ and O$_2$, and also free atoms. The medium $T_{\rm CR}$ class includes the $>40$, $>50$\,K, and probably $>60$\,K regimes. These are best at evaporating CO and CH$_4$. The high-temperature WGH class includes the $>70$ and $>80$\,K (and, in some cases, $>60$\,K) regimes that may promote processes with activation energies of up to about 4000\,K but generally induce only minor changes to cloud chemistry. The $>90$\,K regime has a low $f_T$ and the calculated abundances in most cases are identical to the `model zero'.

We remind that the studied WGH regimes represent a continuous spectra of WGH temperatures and all the $T_{\rm CR}$ regimes simultaneously affect the composition of interstellar clouds. Nevertheless, we discuss them separately in order to identify key chemical processes induced by each $T_{\rm CR}$ class. Their understanding will help explaining the cumulative effects of two or more WGH regimes acting simultaneously, to be studied in a future paper. Evaporation has the dominant effect among the WGH-induced processes on the abundances of chemical species.

\subsection{Major oxygen species}\label{res-o}

\begin{figure*}
 \hspace{-1.0cm}
	\includegraphics{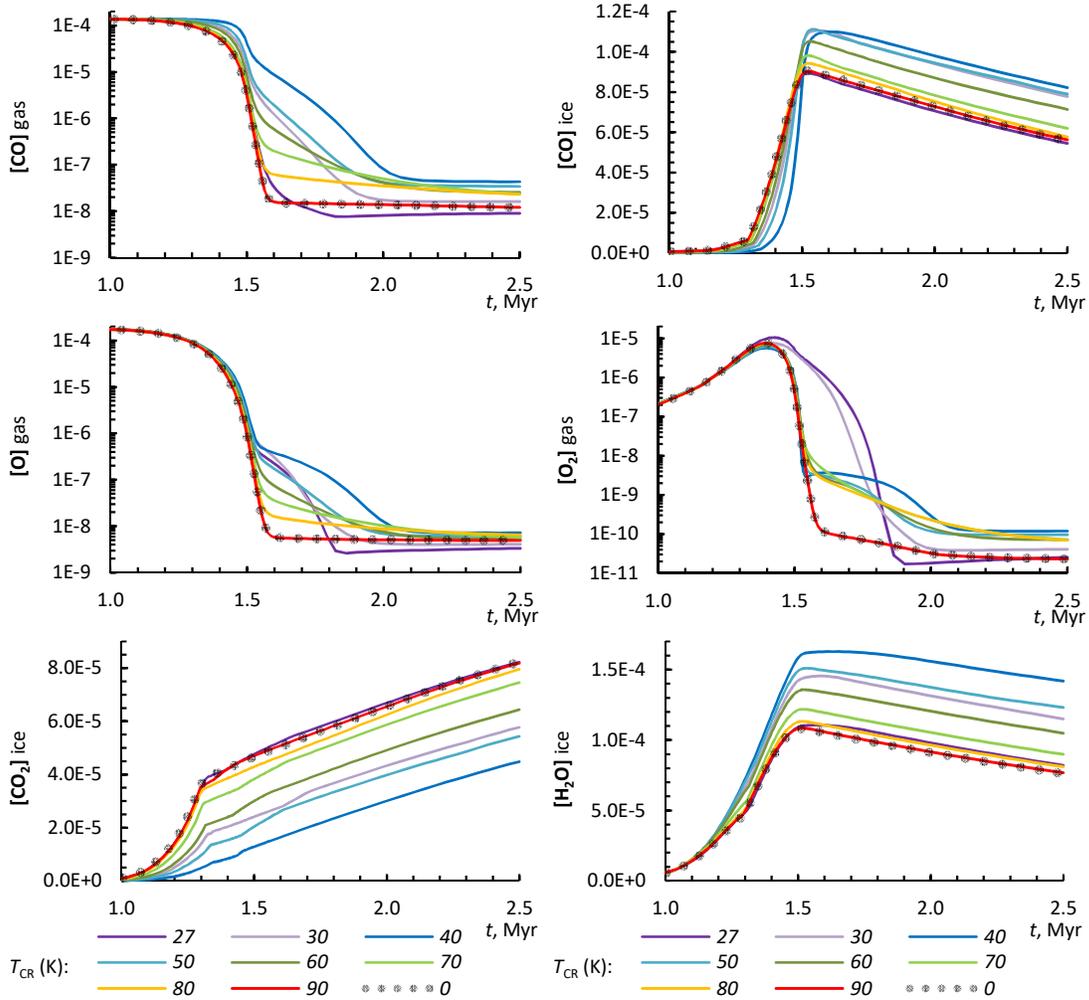}
 \vspace{-15.0cm}
    \caption{Calculated abundances of major oxygen species (including carbon oxides) in gas and ices from simulations with different $T_{\rm CR}$ regimes. Note that gas-phase abundances are displayed with a logarithmic scale, while those of icy species -- with a linear scale. The dots indicate results for `model zero' -- a case with no WGH-induced processes.}
    \label{att-o}
\end{figure*}
Oxygen species with [X] higher than $10^{-5}$ at some point in the simulation are carbon monoxide CO in gas and ice, carbon dioxide CO$_2$ ice, water H$_2$O ice and gaseous atomic oxygen O. Slightly below this threshold is gaseous O$_2$. A quantitative comparison for the calculated abundances of these six species for simulations with all WGH regimes is presented in the appendix tables \ref{tab-co}, \ref{tab-o2} and \ref{tab-h2o}. Figure~\ref{att-o} shows the calculated abundances for simulations considering different WGH regimes.

O, H$_2$O, CO and CO$_2$ form a closely connected gas-ice chemical system. During the first stage of the simulation, gaseous atomic oxygen is transformed to water in the solid phase. The main compounds of carbon are CO in the gas phase and CO and CO$_2$ in ice. For the temperatures and time-scales characteristic to WGH, H$_2$O and CO$_2$ are refractory species and CO is volatile. CO$_2$ is synthesized only on grain surfaces, usually when CO combines with O or OH. The latter two species are accreted from gas or arise \textit{in situ} from the photodissociation of water ice.

The \textbf{medium}-$T_{\rm CR}$ WGH regimes delay the freeze-out of CO for several hundreds of kyr. During this period, the CR-induced dissociation of the CO molecule ensures elevated abundance of C, O and species that are formed from these atoms, including O$_2$, CH$_4$ and carbon chains (Section~\ref{res-cc}). Because CO is adsorbed slowly, oxygen preferably accumulates as water ice. CO$_2$ ice also forms less because of lack of surface CO. Tables \ref{tab-co}, \ref{tab-o2} and \ref{tab-h2o} demonstrate that the $>40$\,K WGH regime is most efficient at inducing these changes.

For the \textbf{low-temperature} WGH regimes, heating to the 30\,K $T_{\rm CR}$ threshold is able to induce some desorption of CO, while heating to 27\,K threshold is not. Both regimes are able to effectively evaporate O$_2$. The low-$T_{\rm CR}$ WGH class is also efficient at evaporating some of O atoms accreted on surfaces. In the gas, these O atoms may combine into O$_2$ (and other gaseous species), instead of the refractory water on the surface. Because of this, the highest abundance of O$_2$ in the gas is reached in the simulations with 27 and 30\,K $T_{\rm CR}$ thresholds.

The \textbf{high}-$T_{\rm CR}$ regimes are not able to induce drastic changes in oxygen chemistry. The $>70$\,K and $>80$\,K simulations have the abundance of gaseous CO increased by a factor of few during the quiescent phase (Table~\ref{tab-co}), but little in absolute numbers. The abundance of O$_2$ during the quiescent stage can be increased by up to an order of magnitude above the $\rm [O_2]\approx4\times10^{-11}$ value of the `model zero'.

\begin{figure*}
 \hspace{-1.0cm}
	\includegraphics{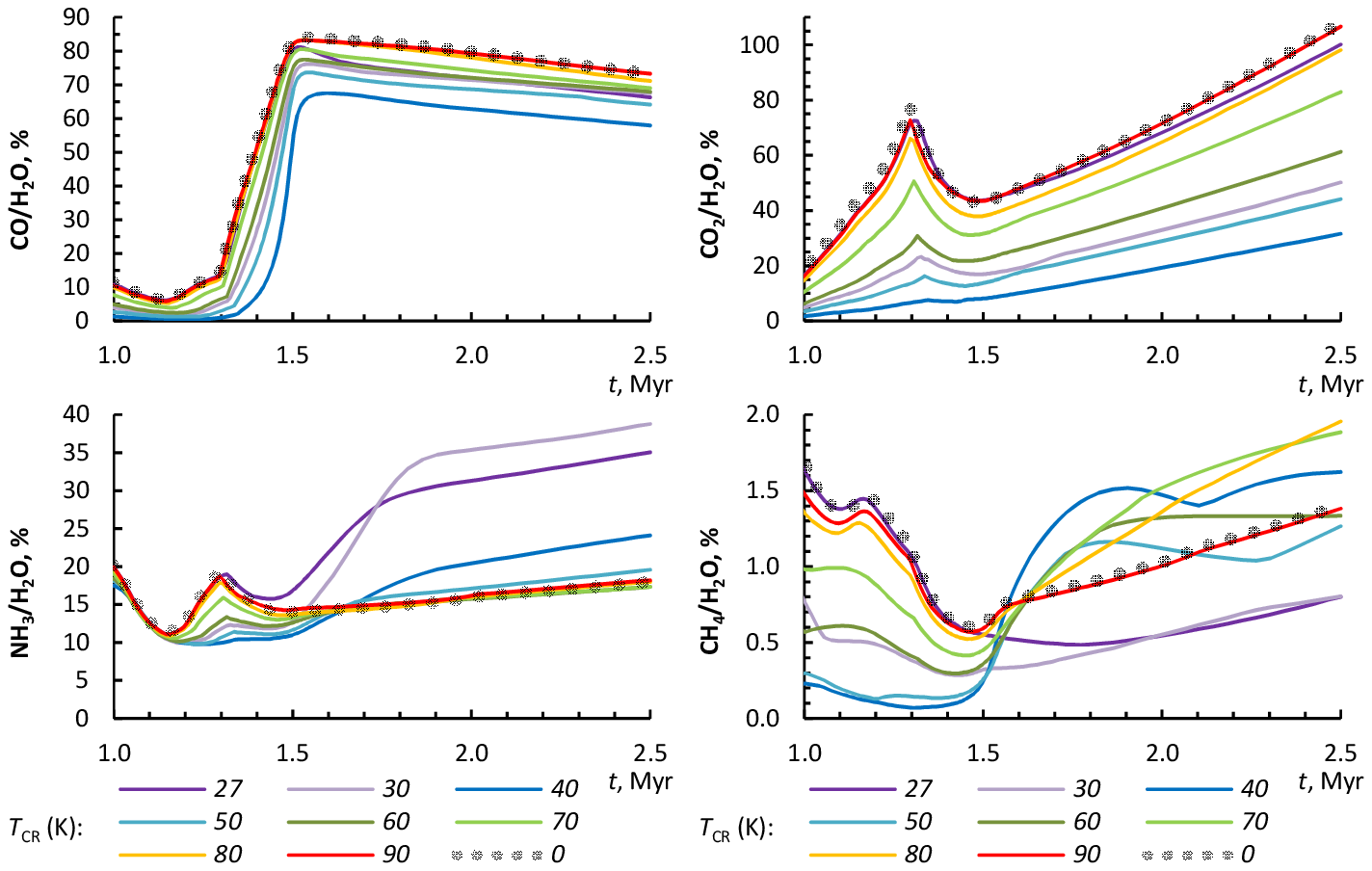}
 \vspace{-19.0cm}
    \caption{Calculated abundances for important icy species, percentage relative to water ice.}
    \label{att-ratio}
\end{figure*}

Observationally, WGH-induced desorption can be constrained by analysing the measured proportions of major ice components at different extinctions. The observed abundance ratios  of icy CO versus icy H$_2$O in the interstellar medium are within the range 8--46 per cent \citep{Teixeira99,Bergin05,Whittet07,Whittet11,Whittet13,Boogert11}. All simulations produce [CO]/[H$_2$O](ice) within this range but the results with medium-temperature WGH thresholds are the closest. For the ratio [CO$_2$]/[H$_2$O](ice) observations indicate range 17-44 per cent, which best agrees with simulations with $T_{\rm CR}$ thresholds in the range $>50$ to $>80$\,K, or even the `model zero' (the same references). Figure~\ref{att-ratio} shows the calculated abundances of CO and CO$_2$ ices relative to water ice.
%

\subsection{Major nitrogen species}\label{res-n}

\begin{figure*}
 \hspace{-1.0cm}
	\includegraphics{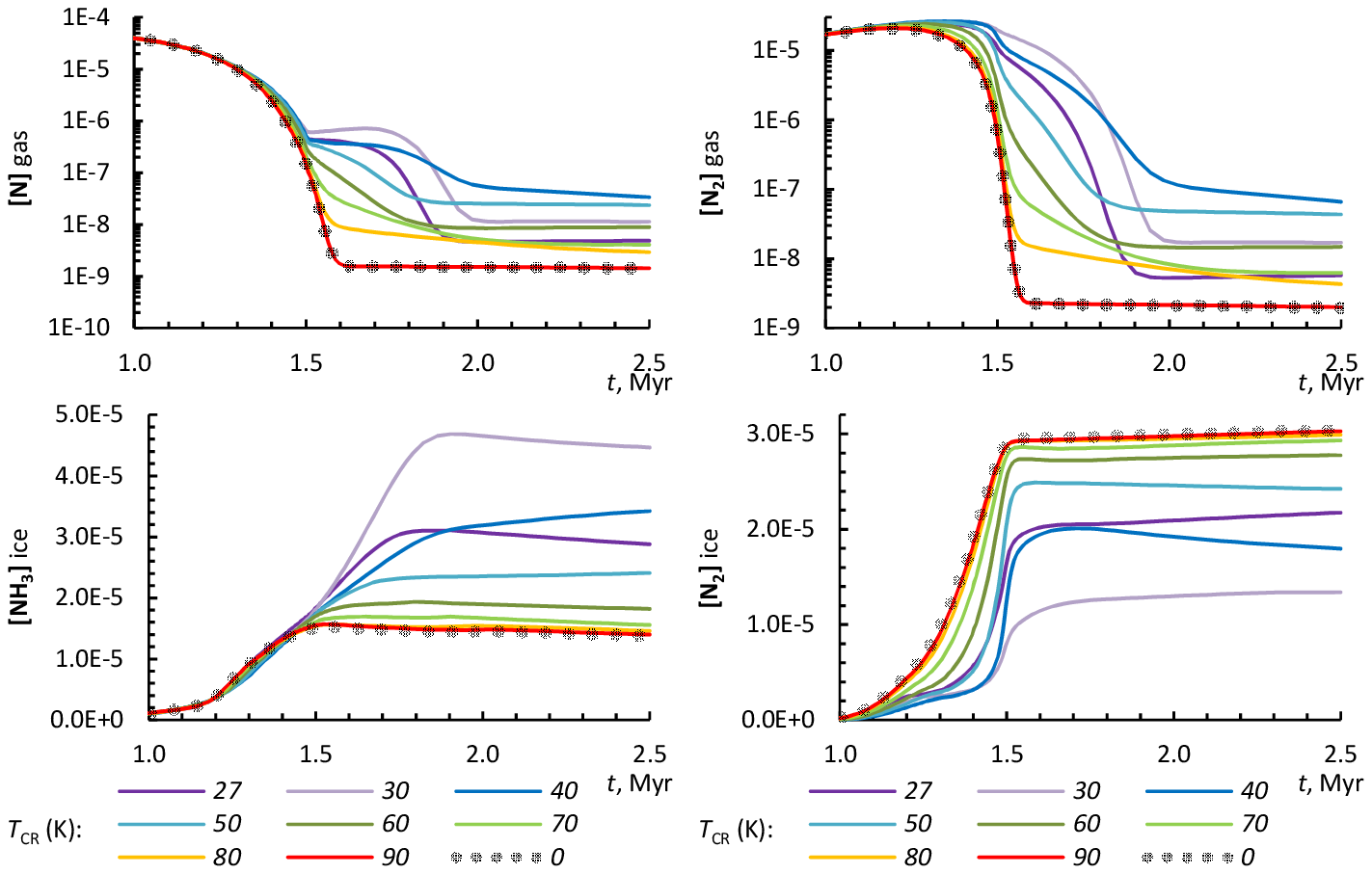}
 \vspace{-19.0cm}
    \caption{Calculated abundances of major nitrogen species in gas and ices from simulations with different $T_{\rm CR}$ regimes. The dots indicate results for the simulation with no WGH-induced processes.}
    \label{att-n}
\end{figure*}
Nitrogen compounds that exceed the $10^{-5}$ abundance limit are N$_2$ (gas and ice) and NH$_3$ (ice), while atomic gaseous N dominates during the early phase of the simulations. Figure~\ref{att-n} shows the calculated abundances of these species, while appendix tables \ref{tab-n} and \ref{tab-nh3} offer a quantitative comparison, to assess the effectiveness of simulations with different $T_{\rm CR}$ thresholds in transforming the chemistry of nitrogen. These data indicate that the simulations with the 30\,K and 40\,K WGH threshold again has the most notable effect on the abundances on N species, when compared to `model zero'.

CR-induced desorption of N$_2$ is the primary means how WGH affects nitrogen chemistry in the model. The other major icy nitrogen molecule, NH$_3$, is refractory at the WGH temperatures. The \textbf{low}-$T_{\rm CR}$ WGH class is able to delay the progress of the freeze-out of N$_2$ by about 300\,kyr. The $>30$\,K regime induces most significant changes in nitrogen chemistry during this freeze-out period. The high abundance of molecular nitrogen also ensures an abundance for atomic N that can be up to three orders of magnitude higher than that of the `model zero'. Gaseous N is continuously depleted to the ices in more refractory forms of nitrogen, such as NH$_3$ and HCN.

The \textbf{medium}-$T_{\rm CR}$ class $>40$\,K and $>50$\,K regimes are similarly effective at evaporating N$_2$. After the freeze-out period, which lasts up to $t\approx1.8$\,Myr, gaseous $\rm [N_2]\approx10^{-7}$, higher than in simulations with other WGH classes  and is decreasing till the end of the simulations. The highest increase for gaseous [N$_2$] for the $>40$\,K simulation is $\approx3000$ times relative to the abundance calculated with `model zero' at $t=1.58$\,Myr.

The medium WGH class maintain the highest gaseous N$_2$ abundances during the quiescent stage -- the residual gas N$_2$ abundance at the end of the simulation is higher by a factor of few tens. This is primarily thanks to cosmic-ray induced diffusion between the layers of ice (Section~\ref{mdl-ch}), which brings some of the subsurface N$_2$ molecules to the surface, where they are available for desorption. Such a diffusion has a higher energy barrier than evaporation and is not so effective for the low-$T_{\rm CR}$ regimes. The effects of such diffusion show only during the quiescent phase but not in the freeze-out period, when the ice layer experiences rapid growth.

Simulations with the \textbf{high}-$T_{\rm CR}$ WGH regimes (here, 60, 70 and 80\,K thresholds) show that these are rather inefficient at delaying the accretion of N$_2$ on grain surfaces, the abundance of gaseous N$_2$ is increased only by a factor of few (Table~\ref{tab-n}).

The introduction of WGH has diverse effects on the abundance of solid ammonia, which is the most abundant nitrogen refractory species. During the initial ice accumulation phase up to $t\approx1.4$\,Myr, WGH reduces [NH$_3$](ice) by a factor of $>$4/5 because less ammonia is produced via surface dissociation of N$_2$. This leads to a notable decrease of (from 19 down to 10 per cent) for the [NH$_3$]/[H$_2$O](ice) ratio because [H$_2$O](ice) is increased at the same time (Section~\ref{res-n}). These variations are most pronounced for the 40\,K WGH regime and are within the 2--23 per cent interval for [NH$_3$]/[H$_2$O](ice) observed in starless cloud cores \citep[cf. Figure~\ref{att-ratio}]{Boogert11,Boogert13}. After 1.4\,Myr, [NH$_3$](ice) is increased at the expense of N$_2$ ice. This is especially true for the low-$T_{\rm CR}$ regimes during the quiescent stage, when [NH$_3$]/[H$_2$O](ice) exceeds 35 per cent. Interestingly, for the low- and medium-$T_{\rm CR}$ regimes, N$_2$ is the last major molecule to retain significant gas-phase abundance. As nitrogen is accumulated on grain surfaces in the form of NH$_3$, the latter molecule covers up to 64 per cent of grain surface.

The abundance of gaseous NH$_3$ remains within the observational data range of $10^{-9}...10^{-7}$ \citep[e.g.,][]{Wirstrom10,Persson12,LeGal14} for the whole duration of the simulations. WGH enhances the relative abundance of gaseous ammonia during the quiescent stage from about $10^{-9}$ in the `case zero' simulation to $10^{-8}$ in simulations with 27, 30, 40 and 50\,K WGH thresholds. Here, gaseous ammonia is replenished by CR-induced photodesorption, efficient thanks to NH$_3$ abundance in the outer surface layer of the icy mantles.

\subsection{Methane and carbon chains}\label{res-cc}

\begin{figure*}
 \hspace{-1.0cm}
	\includegraphics{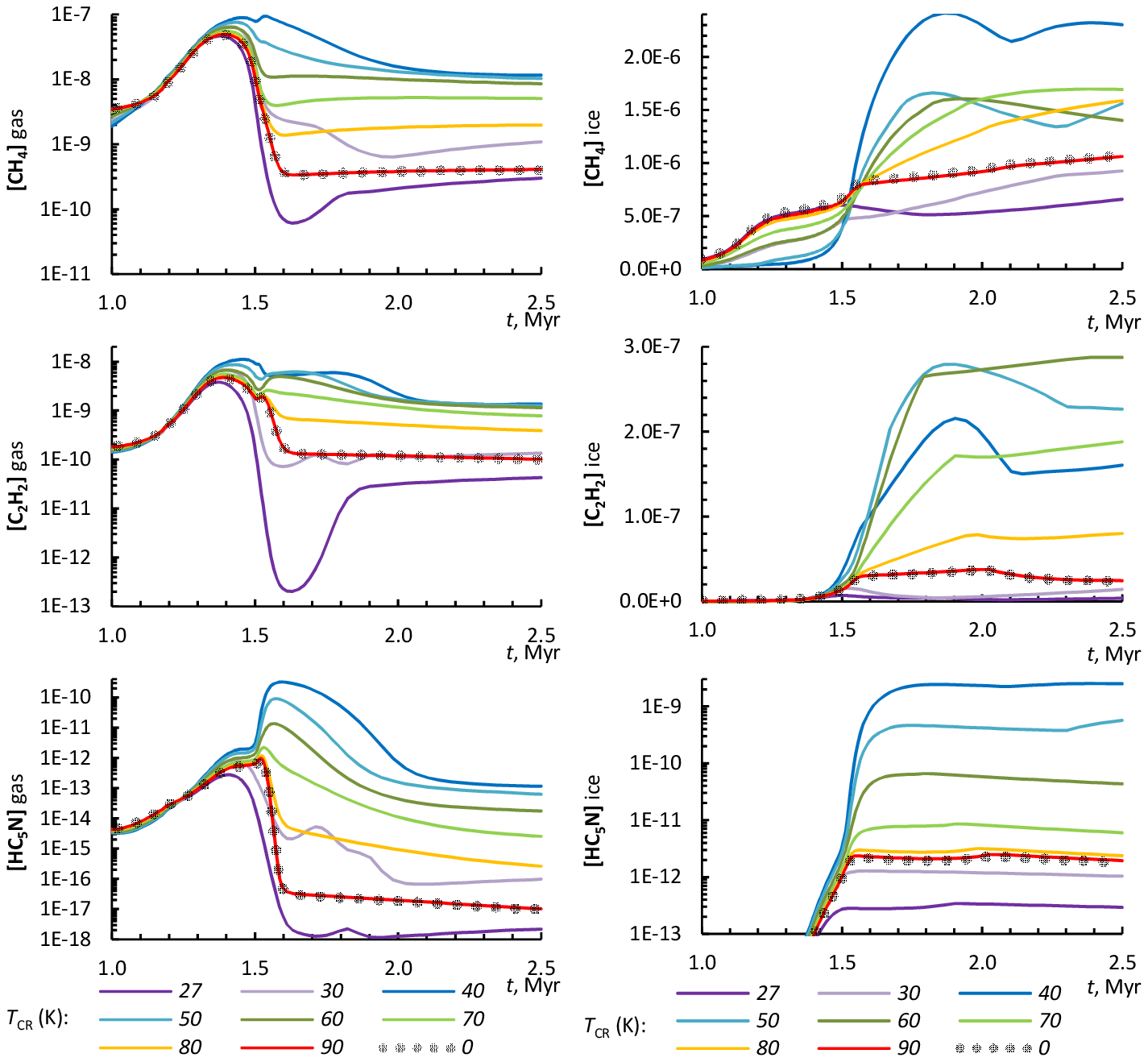}
 \vspace{-15.0cm}
    \caption{Calculated abundances of methane and example carbon chain species in gas and ices from simulations with different $T_{\rm CR}$ regimes. The dots indicate results for a case with no WGH-induced processes.}
    \label{att-cc}
\end{figure*}
Methane and species containing the C-C bond are efficiently produced in the gas; freeze-out reduces their gas-phase abundance. Therefore, it can be expected that WGH-induced desorption will increase the gas-phase abundance of these species. In addition, the abundance of the initial source species -- gas-phase atomic carbon -- depends on CO, which, as a volatile molecule, is strongly affected by WGH-induced evaporation.

These general rules are fulfilled with an exception. Figure~\ref{att-cc} shows that the \textbf{low}-temperature WGH regimes may result in lower abundances of methane or carbon chains during the quiescent phase. For example, [CH$_4$](gas) in the no-WGH simulation is stable at around $\approx4\times10^{-10}$. This value is changed in the $T_{\rm CR}>27$\,K simulation by a factor as low as 0.2. The reduction of gas phase abundance is accompanied with a less pronounced (factor down to 0.6) reduction in abundance of methane ice. More complex species, e.g., C$_2$H$_2$ experience larger deviations and similar abundance changes also for the $>30$\,K regime.

Such a behaviour happens because after the freeze-out of CO, the abundance of C atoms in the gas phase is small, and carbon chains are produced mainly via surface reactions. These are made inefficient by the frequent low-$T_{\rm CR}$ grain heating that causes evaporation of adsorbed C atoms before they can circulate the surface of the grain and combine into molecules.

A similar process occurs also for other WGH regimes but is overshadowed by the effects arising from desorption of CO and other carbon molecules. The \textbf{medium} and \textbf{high} WGH temperature classes ($T_{\rm CR}>40$ to $T_{\rm CR}>80$\,K) result in methane and carbon chain gas-phase abundances that are order(s) of magnitude higher than those of the `model zero' without WGH. This is because the elevated abundance of CO in the gas and in the surface layer serves as a source for the release of C atoms via dissociation. These atoms are then involved in gas and surface synthesis of the organics. The above is a classic example, how a process that seemingly involves only major species (CO) has a direct effect on the abundances and observability of minor interstellar species. Therefore, the WGH regime (that with $T_{\rm CR}>40$\,K) having the most notable effect on CO desorption also most significantly affects the chemistry of carbon chains.

The observed abundance of interstellar methane ice relative to water ice is 0.3--4 per cent \citep{Boogert97}. This agrees best with the simulations with 70 and 80\,K WGH thresholds with [CH$_4$]/[H$_2$O](ice) up to 2 per cent, although other simulations can also produce values above 0.3 per cent.

\subsection{Complex organic molecules}\label{res-com}
\begin{figure*}
 \hspace{-1.0cm}
	\includegraphics{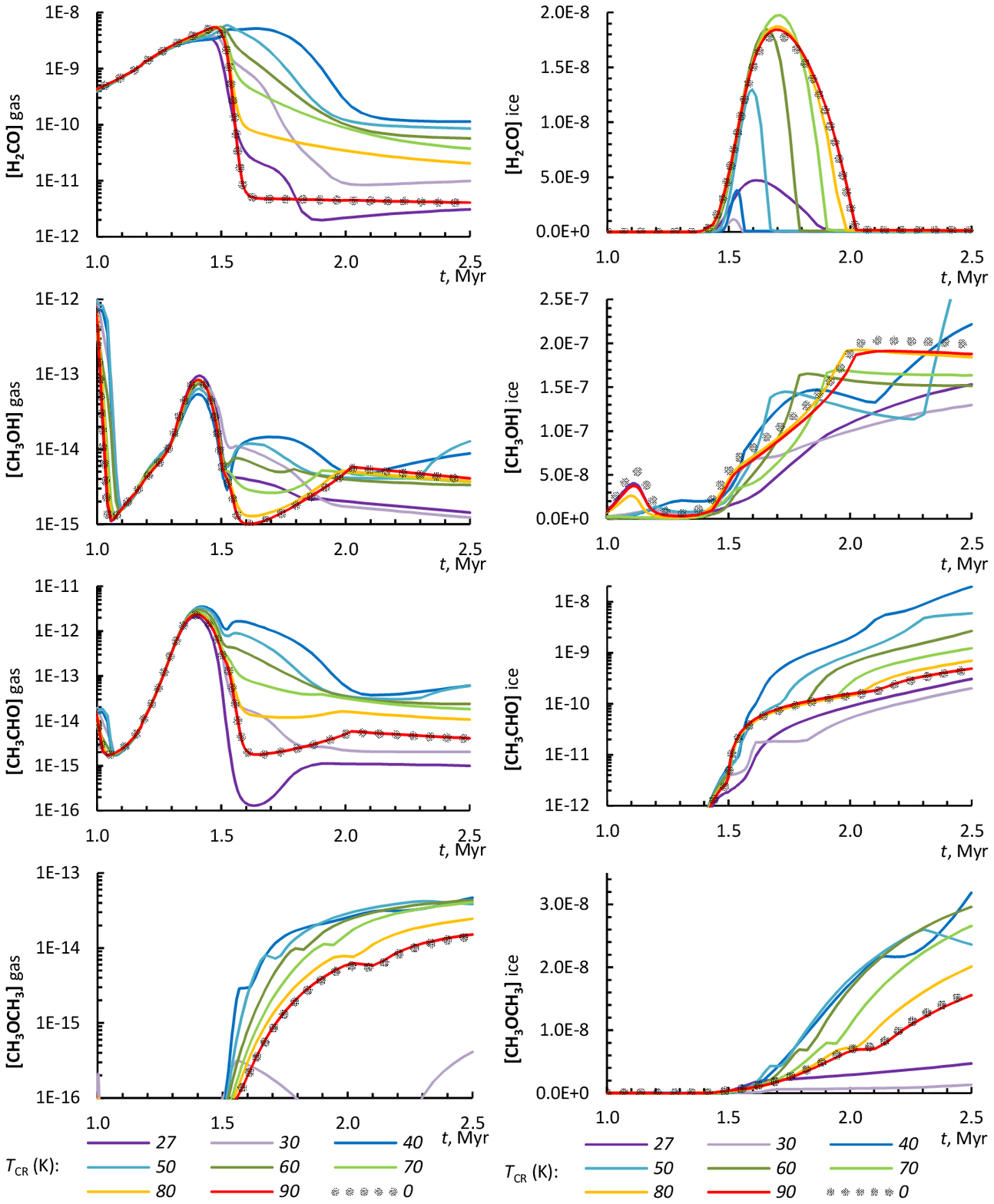}
 \vspace{-10.5cm}
    \caption{Calculated abundances of selected COMs in gas and ices from simulations with different $T_{\rm CR}$ regimes. The dots indicate results for `model zero' with no WGH-induced processes.}
    \label{att-com}
\end{figure*}
WGH affects the abundances of COMs in two ways. First, WGH evaporates volatile organics (e.g., COM precursor formaldehyde, see Figure~\ref{att-com}), increasing their abundance in the gas phase and hampering their accumulation in ices. Second, WGH changes the rate with which CO ice is accumulated. Subsequent processing of CO-rich ices is a major source of COMs \citep[e.g.][]{Garrod06}. Depending on what species (H$_2$O, CH$_4$, NH$_3$, etc.) are co-deposited with CO, some organic species can be preferred over others \citep{K15apj2}.

Figure~\ref{att-com} shows an example of the latter effect -- acetaldehyde CH$_3$CHO. This molecule is synthesized via the reaction $\rm CH_3+HCO$. While the latter is a direct daughter species of CO, the former (CH$_3$) arises from methane or methanol photodissociation. It happens that for the $>40$\,K WGH regime, the accretion of CO is delayed, as well as the surface formation of CH$_4$, which has a similar desorption energy to CO (cf. top panels of Figures \ref{att-o} and \ref{att-cc}). When the density of the core has reached maximum, these two species accumulate simultaneously, creating an upper ice layer, where their photoprocessing products may interact. This results in conditions that are suited for bulk-ice synthesis of CH$_3$CHO. For other WGH regimes, such conditions occur to a lesser extent because CO is accreted earlier and the CO-rich layers have more refractory molecules that change the chemical processes. For example, an admixture of H$_2$O molecules promotes the transformation of CO into the inert CO$_2$ molecules, instead of organics.

The synthesis of methanol CH$_3$OH primarily occurs on grain surfaces and directly depends on CO adsorption. Because WGH generally delays the freeze-out of CO ice, the calculated abundances of CH$_3$OH ice are initially lower than those in the case without WGH for all $T_{\rm CR}$. Because the accretion of water is not delayed, CO and H$_2$O are less mixed. This hampers the production of methanol in bulk ice via the combination of CO with atomic H that arises from the photodissociation of H$_2$O ice.

Despite the effective separation of CO and H$_2$O for the $>50$\,K regimes, [CH$_3$OH](ice) increases to about $4.6\times10^{-7}$ (by a factor of three) during the last two hundred kyr (Figure~\ref{att-com}). This increase occurs solely in the bulk-ice layer that is closest to outer ice surface. To a lesser extent, the $>40$\,K simulation shows a similar phenomenon. Such an increase happens because of a combination of reasons. First, both simulations have characteristic low abundances of hydrogen peroxide because of lack of its parent species -- O$_2$ ice. ([O$_2$](ice) is low because O$_2$ is being kept in the gas phase by the efficient WGH-induced desorption in these simulations.) H$_2$O$_2$ in ice is one of the major consumers of atomic H, photo-produced from H$_2$O. When the supply of H$_2$O$_2$ (and NH$_2$, another sink of H that is continuously replenished from ammonia) is exhausted, the free H atoms are able to hydrogenate CO in a reaction that has a considerable energy barrier, eventually producing methanol. A major sink of H atoms in ice also is their diffusion to the surface (either as H or H$_2$) and evaporation. Such diffusion is facilitated by frequent low-$T_{\rm CR}$ heating that occurs also for the simulation with the 40\,K threshold. For the 50\,K WGH threshold, the diffusion of hydrogen is less efficient, hence H is more involved in chemical reactions, producing the observed increase in the abundance of CH$_3$OH ice.

The explanation above is specific to the present model of interstellar ices and depends on minute details, such as the assumed desorption energy of hydrogen, mobility of light species in ices, frequency of the WGH events and, of course, the assumed $T_{\rm CR}$ threshold. While it may not work for other models, this illustrates the combinations of chemical and physical processes possible in layered interstellar ices.

Dimethyl ether CH$_3$OCH$_3$ is a daughter species of methane and methanol. Its synthesis occurs whenever there is a mix of CH$_4$ and CO ice. As a hydrogenated surface product, methane forms alongside water in the `case zero' simulation. However, the introduction of WGH allows methane to co-adsorb with CO. Like in the case of acetaldehyde, this mixture is responsible for the notable increase in the abundance of dimethyl ether for simulations with WGH. The gas-phase abundance of dimethyl ether relative to hydrogen remains low and achieves only $4\times10^{-14}$ for the 40, 50, 60 and 70\,K WGH thresholds during the quiescent phase.

The observed abundances of gas-phase organic molecules, such as H$_2$CO, HCOOH, HCOOCH$_3$, CH$_3$OCH$_3$, CH$_3$O, CH$_2$CO and CH$_3$CHO, towards molecular cloud cores often are in the range $10^{-12}$--$10^{-9}$, relative to hydrogen. Only that of methanol exceeds $10^{-9}$ \citep[e.g.,][]{Oberg10,Bacmann12,Cernicharo12,Vastel14}. Such a composition might correspond to the proportions of organic molecules in ice, probably indicating a non-selective desorption mechanism at work. The \textsc{Alchemic-Venta} code is fully able to attain and exceed such abundances for these species in ice \citep[see also][]{K15apj2}. However, only volatile species, such as H$_2$CO, or species that can be rapidly synthesized on grain surfaces (HCOOH) attain such abundances in the gas phase during the ice accumulation epoch ($t\approx1.1...1.6$\,Myr). We conclude that WGH-induced desorption is not able to explain the observations of COMs in cold cores.

\section{Conclusions}\label{cncl}

Discussion in Section~\ref{rslt} shows that the chemistry of oxygen, carbon oxides, and nitrogen species is most effectively changed by the $>40$\,K WGH regime. Regimes with higher $T_{\rm CR}$ thresholds mainly induce changes that are similar to the $>40$\,K regime but are less pronounced. The low-temperature $>30$\,K regime is effective at evaporating non-polar small species and introduces specific changes to cloud chemistry, most notably, for nitrogen.

Therefore, we suggest that the $>40$ CR-induced WGH regime should be used in astrochemical modelling with WGH frequency based on updated CR spectra (e.g., as provided in Equation~\ref{wgh3}). This regime has by far the most notable effect among all WGH regimes on the chemistry of cold cores because it is most efficient at evaporating the CO molecule. In the presented model, such heating is also able to induce radial diffusion of volatile species in ices, such as N$_2$, which amplifies the CR-induced desorption effects.

This paper highlights the necessity for a study considering two or more WGH heating regimes acting simultaneously. As seen in the discussions in Section~\ref{rslt}, the chemical consequences of grain heating to even a single WGH temperature can be hardly predictable, with different regimes sometimes producing opposite effects. The physically relevant cumulative effect on the abundances of interstellar species of several WGH regimes therefore needs to be applied and evaluated. If more than one WGH regime is used in an astrochemical model, then the $>30$ and $>50$\,K regimes should be used next, depending also on the aims of the specific research.

From the results of this study we also conclude that the approach on cooling of grains may have to be modified. All WGH regimes, including the 27\,K threshold, induce evaporation of N$_2$ and O$_2$. These molecules are present in the icy mantles of grains in numbers in excess of $10^7$. This is sufficient to control the cooling of grains heated even to 90\,K. Like in the case of CO, only a minority of N$_2$ molecules reside on the outer surface, readily available for evaporative cooling; the time it takes for their diffusion to the surface from the heated ice bulk is another issue.

Finaly, we confirmed that a differentiated molecule accretion in ices facilitate the synthesis of some COMs in bulk-ice \citep[see also][]{K15aa}. The same process may promote a high gas-phase abundance, as in the case of NH$_3$. Both of these `secondary' effects seem to be most pronounced for the $>40$\,K WGH regime.

\section*{Acknowledgements}
The contribution of JK has been funded by ERDF postdoctoral grant No. 1.1.1.2/VIAA/I/16/194 ``Chemical effects of cosmic-ray induced heating of interstellar dust grains''. The contribution of JRK has been supported by ERDF project ``Physical and chemical processes in the interstellar medium'', No 1.1.1.1/16/A/213. Both projects are being implemented in Ventspils University College. We thank to Ventspils City Council for support.

\bibliographystyle{mnras}
\bibliography{crma}

\begin{thebibliography}{}
\makeatletter
\relax
\def\mn@urlcharsother{\let\do\@makeother \do\$\do\&\do\#\do\^\do\_\do\%\do\~}
\def\mn@doi{\begingroup\mn@urlcharsother \@ifnextchar [ {\mn@doi@}
  {\mn@doi@[]}}
\def\mn@doi@[#1]#2{\def\@tempa{#1}\ifx\@tempa\@empty \href
  {http://dx.doi.org/#2} {doi:#2}\else \href {http://dx.doi.org/#2} {#1}\fi
  \endgroup}
\def\mn@eprint#1#2{\mn@eprint@#1:#2::\@nil}
\def\mn@eprint@arXiv#1{\href {http://arxiv.org/abs/#1} {{\tt arXiv:#1}}}
\def\mn@eprint@dblp#1{\href {http://dblp.uni-trier.de/rec/bibtex/#1.xml}
  {dblp:#1}}
\def\mn@eprint@#1:#2:#3:#4\@nil{\def\@tempa {#1}\def\@tempb {#2}\def\@tempc
  {#3}\ifx \@tempc \@empty \let \@tempc \@tempb \let \@tempb \@tempa \fi \ifx
  \@tempb \@empty \def\@tempb {arXiv}\fi \@ifundefined
  {mn@eprint@\@tempb}{\@tempb:\@tempc}{\expandafter \expandafter \csname
  mn@eprint@\@tempb\endcsname \expandafter{\@tempc}}}

\bibitem[\protect\citeauthoryear{{Aannestad} \& {Kenyon}}{{Aannestad} \&
  {Kenyon}}{1979}]{Aannestad79}
{Aannestad} P.~A.,  {Kenyon} S.~J.,  1979, \mn@doi [\apss]
  {10.1007/BF00643496}, \href
  {http://cdsads.u-strasbg.fr/abs/1979Ap%26SS..65..155A} {65, 155}

\bibitem[\protect\citeauthoryear{{Bacmann}, {Taquet}, {Faure}, {Kahane}  \&
  {Ceccarelli}}{{Bacmann} et~al.}{2012}]{Bacmann12}
{Bacmann} A.,  {Taquet} V.,  {Faure} A.,  {Kahane} C.,   {Ceccarelli} C.,
  2012, \mn@doi [\aap] {10.1051/0004-6361/201219207}, \href
  {http://cdsads.u-strasbg.fr/abs/2012A%26A...541L..12B} {541, L12}

\bibitem[\protect\citeauthoryear{{Bergin}, {Melnick}, {Gerakines}, {Neufeld}
  \& {Whittet}}{{Bergin} et~al.}{2005}]{Bergin05}
{Bergin} E.~A.,  {Melnick} G.~J.,  {Gerakines} P.~A.,  {Neufeld} D.~A.,
  {Whittet} D.~C.~B.,  2005, \mn@doi [\apjl] {10.1086/431932}, \href
  {http://cdsads.u-strasbg.fr/abs/2005\apj...627L..33B} {627, L33}

\bibitem[\protect\citeauthoryear{{Boogert}, {Schutte}, {Helmich}, {Tielens}  \&
  {Wooden}}{{Boogert} et~al.}{1997}]{Boogert97}
{Boogert} A.~C.~A.,  {Schutte} W.~A.,  {Helmich} F.~P.,  {Tielens} A.~G.~G.~M.,
    {Wooden} D.~H.,  1997, \aap, \href
  {http://cdsads.u-strasbg.fr/abs/1997A%26A...317..929B} {317, 929}

\bibitem[\protect\citeauthoryear{{Boogert} et~al.,}{{Boogert}
  et~al.}{2011}]{Boogert11}
{Boogert} A.~C.~A.,  et~al., 2011, \mn@doi [\apj] {10.1088/0004-637X/729/2/92},
  \href {http://cdsads.u-strasbg.fr/abs/2011ApJ...729...92B} {729, 92}

\bibitem[\protect\citeauthoryear{{Boogert}, {Chiar}, {Knez}, {{\"O}berg},
  {Mundy}, {Pendleton}, {Tielens}  \& {van Dishoeck}}{{Boogert}
  et~al.}{2013}]{Boogert13}
{Boogert} A.~C.~A.,  {Chiar} J.~E.,  {Knez} C.,  {{\"O}berg} K.~I.,  {Mundy}
  L.~G.,  {Pendleton} Y.~J.,  {Tielens} A.~G.~G.~M.,   {van Dishoeck} E.~F.,
  2013, \mn@doi [ApJ] {10.1088/0004-637X/777/1/73}, \href
  {http://cdsads.u-strasbg.fr/abs/2013ApJ...777...73B} {777, 73}

\bibitem[\protect\citeauthoryear{{Cecchi-Pestellini} \&
  {Aiello}}{{Cecchi-Pestellini} \& {Aiello}}{1992}]{Cecchi92}
{Cecchi-Pestellini} C.,  {Aiello} S.,  1992, \mn@doi [\mnras]
  {10.1093/mnras/258.1.125}, \href
  {http://cdsads.u-strasbg.fr/abs/1992MNRAS.258..125C} {258, 125}

\bibitem[\protect\citeauthoryear{{Cernicharo}, {Marcelino}, {Roueff}, {Gerin},
  {Jim{\'e}nez-Escobar}  \& {Mu{\~n}oz Caro}}{{Cernicharo}
  et~al.}{2012}]{Cernicharo12}
{Cernicharo} J.,  {Marcelino} N.,  {Roueff} E.,  {Gerin} M.,
  {Jim{\'e}nez-Escobar} A.,   {Mu{\~n}oz Caro} G.~M.,  2012, \mn@doi [\apjl]
  {10.1088/2041-8205/759/2/L43}, \href
  {http://cdsads.u-strasbg.fr/abs/2012\apj...759L..43C} {759, L43}

\bibitem[\protect\citeauthoryear{{Chabot}}{{Chabot}}{2016}]{Chabot16}
{Chabot} M.,  2016, \mn@doi [\aap] {10.1051/0004-6361/201425441}, \href
  {http://cdsads.u-strasbg.fr/abs/2016A%26A...585A..15C} {585, A15}

\bibitem[\protect\citeauthoryear{{Cummings} et~al.,}{{Cummings}
  et~al.}{2016}]{Cummings16}
{Cummings} A.~C.,  et~al., 2016, \mn@doi [\apj] {10.3847/0004-637X/831/1/18},
  \href {http://cdsads.u-strasbg.fr/abs/2016ApJ...831...18C} {831, 18}

\bibitem[\protect\citeauthoryear{{Cuppen}, {Morata}  \& {Herbst}}{{Cuppen}
  et~al.}{2006}]{Cuppen06}
{Cuppen} H.~M.,  {Morata} O.,   {Herbst} E.,  2006, \mn@doi [MNRAS]
  {10.1111/j.1365-2966.2006.10079.x}, \href
  {http://cdsads.u-strasbg.fr/abs/2006MNRAS.367.1757C} {367, 1757}

\bibitem[\protect\citeauthoryear{{Draine} \& {Li}}{{Draine} \&
  {Li}}{2001}]{Draine01}
{Draine} B.~T.,  {Li} A.,  2001, \mn@doi [\apj] {10.1086/320227}, \href
  {http://cdsads.u-strasbg.fr/abs/2001ApJ...551..807D} {551, 807}

\bibitem[\protect\citeauthoryear{{Duley}}{{Duley}}{1973}]{Duley73}
{Duley} W.~W.,  1973, \mn@doi [\apss] {10.1007/BF00647650}, \href
  {http://cdsads.u-strasbg.fr/abs/1973Ap%26SS..23...43D} {23, 43}

\bibitem[\protect\citeauthoryear{{Garrod} \& {Herbst}}{{Garrod} \&
  {Herbst}}{2006}]{Garrod06}
{Garrod} R.~T.,  {Herbst} E.,  2006, \mn@doi [\aap]
  {10.1051/0004-6361:20065560}, \href
  {http://cdsads.u-strasbg.fr/abs/2006A%26A...457..927G} {457, 927}

\bibitem[\protect\citeauthoryear{{Garrod}, {Weaver}  \& {Herbst}}{{Garrod}
  et~al.}{2008}]{Garrod08}
{Garrod} R.~T.,  {Weaver} S.~L.~W.,   {Herbst} E.,  2008, \mn@doi [\apj]
  {10.1086/588035}, \href {http://cdsads.u-strasbg.fr/abs/2008\apj...682..283G}
  {682, 283}

\bibitem[\protect\citeauthoryear{{Greenberg} \& {Hong}}{{Greenberg} \&
  {Hong}}{1974}]{Greenberg74}
{Greenberg} J.~M.,  {Hong} S.-S.,  1974, in {Kerr} F.~J.,  {Simonson} S.~C.,
  eds,  IAU Symposium Vol. 60, Galactic Radio Astronomy. pp 155--177

\bibitem[\protect\citeauthoryear{{Greenberg} \& {Yencha}}{{Greenberg} \&
  {Yencha}}{1973}]{Greenberg73}
{Greenberg} J.~M.,  {Yencha} A.~J.,  1973, in {Greenberg} J.~M.,  {van de
  Hulst} H.~C.,  eds,  IAU Symposium Vol. 52, Interstellar Dust and Related
  Topics. p.~369

\bibitem[\protect\citeauthoryear{{Hasegawa} \& {Herbst}}{{Hasegawa} \&
  {Herbst}}{1993}]{Hasegawa93}
{Hasegawa} T.~I.,  {Herbst} E.,  1993, \mnras, \href
  {http://adsabs.harvard.edu/abs/1993MNRAS.261...83H} {261, 83}

\bibitem[\protect\citeauthoryear{{Hasegawa}, {Herbst}  \& {Leung}}{{Hasegawa}
  et~al.}{1992}]{Hasegawa92}
{Hasegawa} T.~I.,  {Herbst} E.,   {Leung} C.~M.,  1992, \mn@doi [\apjs]
  {10.1086/191713}, \href {http://adsabs.harvard.edu/abs/1992ApJS...82..167H}
  {82, 167}

\bibitem[\protect\citeauthoryear{{Hocuk}, {Sz{\H u}cs}, {Caselli}, {Cazaux},
  {Spaans}  \& {Esplugues}}{{Hocuk} et~al.}{2017}]{Hocuk17}
{Hocuk} S.,  {Sz{\H u}cs} L.,  {Caselli} P.,  {Cazaux} S.,  {Spaans} M.,
  {Esplugues} G.~B.,  2017, \mn@doi [\aap] {10.1051/0004-6361/201629944}, \href
  {http://cdsads.u-strasbg.fr/abs/2017A%26A...604A..58H} {604, A58}

\bibitem[\protect\citeauthoryear{{Indriolo}, {Fields}  \& {McCall}}{{Indriolo}
  et~al.}{2009}]{Indriolo09}
{Indriolo} N.,  {Fields} B.~D.,   {McCall} B.~J.,  2009, \mn@doi [\apj]
  {10.1088/0004-637X/694/1/257}, \href
  {http://cdsads.u-strasbg.fr/abs/2009ApJ...694..257I} {694, 257}

\bibitem[\protect\citeauthoryear{{Iqbal} \& {Wakelam}}{{Iqbal} \&
  {Wakelam}}{2018}]{Iqbal18}
{Iqbal} W.,  {Wakelam} V.,  2018, \mn@doi [\aap] {10.1051/0004-6361/201732486},
  \href {http://cdsads.u-strasbg.fr/abs/2018A%26A...615A..20I} {615, A20}

\bibitem[\protect\citeauthoryear{{Ivlev}, {Padovani}, {Galli}  \&
  {Caselli}}{{Ivlev} et~al.}{2015}]{Ivlev15p}
{Ivlev} A.~V.,  {Padovani} M.,  {Galli} D.,   {Caselli} P.,  2015, \mn@doi
  [\apj] {10.1088/0004-637X/812/2/135}, \href
  {http://cdsads.u-strasbg.fr/abs/2015ApJ...812..135I} {812, 135}

\bibitem[\protect\citeauthoryear{{Kalv{\= a}ns}}{{Kalv{\= a}ns}}{2014}]{K14}
{Kalv{\= a}ns} J.,  2014, Baltic Astronomy, \href
  {http://cdsads.u-strasbg.fr/abs/2014BaltA..23..137K} {23, 137}

\bibitem[\protect\citeauthoryear{{Kalv{\= a}ns}}{{Kalv{\= a}ns}}{2015a}]{K15aa}
{Kalv{\= a}ns} J.,  2015a, \aap, 573, A38

\bibitem[\protect\citeauthoryear{{Kalv{\= a}ns}}{{Kalv{\=
  a}ns}}{2015b}]{K15apj1}
{Kalv{\= a}ns} J.,  2015b, \mn@doi [\apj] {10.1088/0004-637X/803/2/52}, \href
  {http://cdsads.u-strasbg.fr/abs/2015ApJ...803...52K} {803, 52}

\bibitem[\protect\citeauthoryear{{Kalv{\= a}ns}}{{Kalv{\=
  a}ns}}{2015c}]{K15apj2}
{Kalv{\= a}ns} J.,  2015c, \mn@doi [\apj] {10.1088/0004-637X/806/2/196}, \href
  {http://cdsads.u-strasbg.fr/abs/2015ApJ...806..196K} {806, 196}

\bibitem[\protect\citeauthoryear{{Kalv{\= a}ns}}{{Kalv{\= a}ns}}{2016}]{K16}
{Kalv{\= a}ns} J.,  2016, \mn@doi [\apjs] {10.3847/0067-0049/224/2/42}, \href
  {http://cdsads.u-strasbg.fr/abs/2016ApJS..224...42K} {224, 42 (Paper~I)}

\bibitem[\protect\citeauthoryear{{Kalv{\= a}ns}}{{Kalv{\= a}ns}}{2018a}]{K18ap}
{Kalv{\= a}ns} J.,  2018a, \mn@doi [\apjs] {10.3847/1538-4365/aae527}, \href
  {http://cdsads.u-strasbg.fr/abs/2018arXiv181000708Khttp://cdsads.u-strasbg.fr/abs/2018arXiv181000708K}
  {239, 6 (Paper~II)}

\bibitem[\protect\citeauthoryear{{Kalv{\= a}ns}}{{Kalv{\= a}ns}}{2018b}]{K18mn}
{Kalv{\= a}ns} J.,  2018b, \mn@doi [\mnras] {10.1093/mnras/sty1172}, \href
  {http://adsabs.harvard.edu/abs/2018MNRAS.478.2753K} {478, 2753}

\bibitem[\protect\citeauthoryear{{Kalv{\= a}ns}, {Shmeld}, {Kalnin}  \&
  {Hocuk}}{{Kalv{\= a}ns} et~al.}{2017}]{K17}
{Kalv{\= a}ns} J.,  {Shmeld} I.,  {Kalnin} J.~R.,   {Hocuk} S.,  2017, \mn@doi
  [\mnras] {10.1093/mnras/stx174}, \href
  {http://adsabs.harvard.edu/abs/2017MNRAS.467.1763K} {467, 1763}

\bibitem[\protect\citeauthoryear{{Le Gal}, {Hily-Blant}, {Faure}, {Pineau des
  For{\^e}ts}, {Rist}  \& {Maret}}{{Le Gal} et~al.}{2014}]{LeGal14}
{Le Gal} R.,  {Hily-Blant} P.,  {Faure} A.,  {Pineau des For{\^e}ts} G.,
  {Rist} C.,   {Maret} S.,  2014, \mn@doi [\aap] {10.1051/0004-6361/201322386},
  \href {http://cdsads.u-strasbg.fr/abs/2014A%26A...562A..83L} {562, A83}

\bibitem[\protect\citeauthoryear{{Leger}, {Jura}  \& {Omont}}{{Leger}
  et~al.}{1985}]{Leger85}
{Leger} A.,  {Jura} M.,   {Omont} A.,  1985, \aap, \href
  {http://adsabs.harvard.edu/abs/1985A%26A...144..147L} {144, 147}

\bibitem[\protect\citeauthoryear{{Lippok} et~al.,}{{Lippok}
  et~al.}{2013}]{Lippok13}
{Lippok} N.,  et~al., 2013, \mn@doi [\aap] {10.1051/0004-6361/201322129}, \href
  {http://cdsads.u-strasbg.fr/abs/2013A%26A...560A..41L} {560, A41}

\bibitem[\protect\citeauthoryear{{McElroy}, {Walsh}, {Markwick}, {Cordiner},
  {Smith}  \& {Millar}}{{McElroy} et~al.}{2013}]{McElroy13}
{McElroy} D.,  {Walsh} C.,  {Markwick} A.~J.,  {Cordiner} M.~A.,  {Smith} K.,
  {Millar} T.~J.,  2013, \mn@doi [\aap] {10.1051/0004-6361/201220465}, \href
  {http://cdsads.u-strasbg.fr/abs/2013A%26A...550A..36M} {550, A36}

\bibitem[\protect\citeauthoryear{{Morlino}, {Gabici}  \& {Krause}}{{Morlino}
  et~al.}{2015}]{Morlino15}
{Morlino} G.,  {Gabici} S.,   {Krause} J.,  2015, ArXiv e-print 1509.05128,
  \href {http://cdsads.u-strasbg.fr/abs/2015arXiv150905128M} {}

\bibitem[\protect\citeauthoryear{{Moskalenko}, {Strong}, {Ormes}  \&
  {Potgieter}}{{Moskalenko} et~al.}{2002}]{Moskalenko02}
{Moskalenko} I.~V.,  {Strong} A.~W.,  {Ormes} J.~F.,   {Potgieter} M.~S.,
  2002, \mn@doi [\apj] {10.1086/324402}, \href
  {http://cdsads.u-strasbg.fr/abs/2002ApJ...565..280M} {565, 280}

\bibitem[\protect\citeauthoryear{{Nejad}, {Williams}  \& {Charnley}}{{Nejad}
  et~al.}{1990}]{Nejad90}
{Nejad} L.~A.~M.,  {Williams} D.~A.,   {Charnley} S.~B.,  1990, \mnras, \href
  {http://cdsads.u-strasbg.fr/abs/1990\mnras.246..183N} {246, 183}

\bibitem[\protect\citeauthoryear{{{\"O}berg}, {Bottinelli}, {J{\o}rgensen}  \&
  {van Dishoeck}}{{{\"O}berg} et~al.}{2010}]{Oberg10}
{{\"O}berg} K.~I.,  {Bottinelli} S.,  {J{\o}rgensen} J.~K.,   {van Dishoeck}
  E.~F.,  2010, \mn@doi [\apj] {10.1088/0004-637X/716/1/825}, \href
  {http://cdsads.u-strasbg.fr/abs/2010ApJ...716..825O} {716, 825}

\bibitem[\protect\citeauthoryear{{Padovani}, {Galli}  \&
  {Glassgold}}{{Padovani} et~al.}{2009}]{Padovani09}
{Padovani} M.,  {Galli} D.,   {Glassgold} A.~E.,  2009, \mn@doi [\aap]
  {10.1051/0004-6361/200911794}, \href
  {http://cdsads.u-strasbg.fr/abs/2009A%26A...501..619P} {501, 619}

\bibitem[\protect\citeauthoryear{{Persson} et~al.,}{{Persson}
  et~al.}{2012}]{Persson12}
{Persson} C.~M.,  et~al., 2012, \mn@doi [\aap] {10.1051/0004-6361/201118686},
  \href {http://cdsads.u-strasbg.fr/abs/2012A%26A...543A.145P} {543, A145}

\bibitem[\protect\citeauthoryear{{Reboussin}, {Wakelam}, {Guilloteau}  \&
  {Hersant}}{{Reboussin} et~al.}{2014}]{Reboussin14}
{Reboussin} L.,  {Wakelam} V.,  {Guilloteau} S.,   {Hersant} F.,  2014, \mn@doi
  [\mnras] {10.1093/mnras/stu462}, \href
  {http://cdsads.u-strasbg.fr/abs/2014MNRAS.440.3557R} {440, 3557}

\bibitem[\protect\citeauthoryear{{Roberts}, {Rawlings}, {Viti}  \&
  {Williams}}{{Roberts} et~al.}{2007}]{Roberts07}
{Roberts} J.~F.,  {Rawlings} J.~M.~C.,  {Viti} S.,   {Williams} D.~A.,  2007,
  \mn@doi [\mnras] {10.1111/j.1365-2966.2007.12402.x}, \href
  {http://adsabs.harvard.edu/abs/2007MNRAS.382..733R} {382, 733}

\bibitem[\protect\citeauthoryear{{Semenov} et~al.,}{{Semenov}
  et~al.}{2010}]{Semenov10}
{Semenov} D.,  et~al., 2010, \mn@doi [\aap] {10.1051/0004-6361/201015149},
  \href {http://cdsads.u-strasbg.fr/abs/2010A%26A...522A..42S} {522, A42}

\bibitem[\protect\citeauthoryear{{Shen}, {Greenberg}, {Schutte}  \& {van
  Dishoeck}}{{Shen} et~al.}{2004}]{Shen04}
{Shen} C.~J.,  {Greenberg} J.~M.,  {Schutte} W.~A.,   {van Dishoeck} E.~F.,
  2004, \mn@doi [\aap] {10.1051/0004-6361:20031669}, \href
  {http://cdsads.u-strasbg.fr/abs/2004A%26A...415..203S} {415, 203}

\bibitem[\protect\citeauthoryear{{Tabak}}{{Tabak}}{1987}]{Tabak87}
{Tabak} R.~G.,  1987, \mn@doi [ApSS] {10.1007/BF00636461}, \href
  {http://cdsads.u-strasbg.fr/abs/1987Ap%26SS.134..145T} {134, 145}

\bibitem[\protect\citeauthoryear{{Teixeira} \& {Emerson}}{{Teixeira} \&
  {Emerson}}{1999}]{Teixeira99}
{Teixeira} T.~C.,  {Emerson} J.~P.,  1999, \aap, \href
  {http://cdsads.u-strasbg.fr/abs/1999A%26A...351..292T} {351, 292}

\bibitem[\protect\citeauthoryear{{Thi}, {Woitke}  \& {Kamp}}{{Thi}
  et~al.}{2010}]{Thi10}
{Thi} W.-F.,  {Woitke} P.,   {Kamp} I.,  2010, \mn@doi [\mnras]
  {10.1111/j.1365-2966.2009.16162.x}, \href
  {http://cdsads.u-strasbg.fr/abs/2010MNRAS.407..232T} {407, 232}

\bibitem[\protect\citeauthoryear{{Vastel}, {Ceccarelli}, {Lefloch}  \&
  {Bachiller}}{{Vastel} et~al.}{2014}]{Vastel14}
{Vastel} C.,  {Ceccarelli} C.,  {Lefloch} B.,   {Bachiller} R.,  2014, \mn@doi
  [\apjl] {10.1088/2041-8205/795/1/L2}, \href
  {http://cdsads.u-strasbg.fr/abs/2014ApJ...795L...2V} {795, L2}

\bibitem[\protect\citeauthoryear{{Vasyunin}, {Caselli}, {Dulieu}  \&
  {Jim{\'e}nez-Serra}}{{Vasyunin} et~al.}{2017}]{Vasyunin17}
{Vasyunin} A.~I.,  {Caselli} P.,  {Dulieu} F.,   {Jim{\'e}nez-Serra} I.,  2017,
  \mn@doi [\apj] {10.3847/1538-4357/aa72ec}, \href
  {http://cdsads.u-strasbg.fr/abs/2017ApJ...842...33V} {842, 33}

\bibitem[\protect\citeauthoryear{{Whittet}, {Shenoy}, {Bergin}, {Chiar},
  {Gerakines}, {Gibb}, {Melnick}  \& {Neufeld}}{{Whittet}
  et~al.}{2007}]{Whittet07}
{Whittet} D.~C.~B.,  {Shenoy} S.~S.,  {Bergin} E.~A.,  {Chiar} J.~E.,
  {Gerakines} P.~A.,  {Gibb} E.~L.,  {Melnick} G.~J.,   {Neufeld} D.~A.,  2007,
  \mn@doi [\apj] {10.1086/509772}, \href
  {http://cdsads.u-strasbg.fr/abs/2007\apj...655..332W} {655, 332}

\bibitem[\protect\citeauthoryear{{Whittet}, {Cook}, {Herbst}, {Chiar}  \&
  {Shenoy}}{{Whittet} et~al.}{2011}]{Whittet11}
{Whittet} D.~C.~B.,  {Cook} A.~M.,  {Herbst} E.,  {Chiar} J.~E.,   {Shenoy}
  S.~S.,  2011, \mn@doi [\apj] {10.1088/0004-637X/742/1/28}, \href
  {http://cdsads.u-strasbg.fr/abs/2011ApJ...742...28W} {742, 28}

\bibitem[\protect\citeauthoryear{{Whittet}, {Poteet}, {Chiar}, {Pagani},
  {Bajaj}, {Horne}, {Shenoy}  \& {Adamson}}{{Whittet} et~al.}{2013}]{Whittet13}
{Whittet} D.~C.~B.,  {Poteet} C.~A.,  {Chiar} J.~E.,  {Pagani} L.,  {Bajaj}
  V.~M.,  {Horne} D.,  {Shenoy} S.~S.,   {Adamson} A.~J.,  2013, \mn@doi [\apj]
  {10.1088/0004-637X/774/2/102}, \href
  {http://cdsads.u-strasbg.fr/abs/2013ApJ...774..102W} {774, 102}

\bibitem[\protect\citeauthoryear{{Wirstr{\"o}m} et~al.,}{{Wirstr{\"o}m}
  et~al.}{2010}]{Wirstrom10}
{Wirstr{\"o}m} E.~S.,  et~al., 2010, \mn@doi [\aap]
  {10.1051/0004-6361/200913766}, \href
  {http://cdsads.u-strasbg.fr/abs/2010A%26A...522A..19W} {522, A19}

\bibitem[\protect\citeauthoryear{{d'Hendecourt}, {Allamandola}, {Baas}  \&
  {Greenberg}}{{d'Hendecourt} et~al.}{1982}]{dHendecourt82}
{d'Hendecourt} L.~B.,  {Allamandola} L.~J.,  {Baas} F.,   {Greenberg} J.~M.,
  1982, \aap, \href {http://adsabs.harvard.edu/abs/1982A%26A...109L..12D} {109,
  L12}

\bibitem[\protect\citeauthoryear{{de Barros}, {Domaracka}, {Andrade}, {Boduch},
  {Rothard}  \& {da Silveira}}{{de Barros} et~al.}{2011}]{deBarros11}
{de Barros} A.~L.~F.,  {Domaracka} A.,  {Andrade} D.~P.~P.,  {Boduch} P.,
  {Rothard} H.,   {da Silveira} E.~F.,  2011, \mn@doi [\mnras]
  {10.1111/j.1365-2966.2011.19587.x}, \href
  {http://cdsads.u-strasbg.fr/abs/2011MNRAS.418.1363D} {418, 1363}

\makeatother
\end{thebibliography}

\appendix

\section{Comparison tables to assess the effectiveness of different WGH regimes} \label{app-tab}

This appendix presents data calculation results -- gas and icy species' abundances at different WGH regimes -- that are quantitatively compared to the results of the `model zero' with no WGH-induced processes at all. The comparison is done for two characteristic points in time -- at $t=1.5$\,Myr, when cloud contraction stops, and at $t=2.5$\,Myr, the end time of the modelled quiescent phase.

\begin{table}
	\centering
\caption{Calculated abundance ratio for the CO molecule in gas and ices: results for simulations with WGH (eight $T_{\rm CR}$ thresholds, listed in the table) versus results without WGH. The latter are provided within the table.}
\label{tab-co}
	\begin{tabular}{lcccc}
  \hline
 & \multicolumn{2}{c}{CO gas} & \multicolumn{2}{c}{CO ice} \\
Abundance, & 1.5 Myr & 2.5 Myr & 1.5 Myr & 2.5 Myr \\
no WGH, cm$^{-3}$ & 3.67E-06 & 1.20E-08 & 8.86E-05 & 5.58E-05 \\
\hline
($T_{\rm CR}$, K) & \multicolumn{4}{c}{[CO]/[CO\,no\,WGH]} \\
>27 & 1.20 & 0.75 & 0.98 & 0.98 \\
>30 & 4.04 & 1.35 & 1.13 & 1.39 \\
>40 & 12.54 & 3.59 & 0.91 & 1.47 \\
>50 & 5.47 & 2.84 & 1.12 & 1.41 \\
>60 & 2.59 & 2.12 & 1.12 & 1.28 \\
>70 & 1.54 & 1.98 & 1.07 & 1.11 \\
>80 & 1.15 & 1.92 & 1.04 & 1.03 \\
>90 & 1.02 & 1.01 & 0.99 & 1.01 \\
\hline
	\end{tabular}
\end{table}

\begin{table}
	\centering
\caption{Calculated abundance ratio for the O atom and the O$_2$ molecule in gas phase: results for simulations with WGH (eight $T_{\rm CR}$ thresholds) versus results without WGH. The latter are provided within the table.}
\label{tab-o2}
	\begin{tabular}{lcccc}
  \hline
 & \multicolumn{2}{c}{O gas} & \multicolumn{2}{c}{O$_2$ gas} \\
Abundance, & 1.5 Myr & 2.5 Myr & 1.5 Myr & 2.5 Myr \\
no WGH, cm$^{-3}$ & 1.02E-06 & 4.91E-09 & 5.44E-07 & 2.31E-11 \\
\hline
($T_{\rm CR}$, K) & \multicolumn{2}{c}{[O]/[O,\,no\,WGH]} & \multicolumn{2}{c}{[O$_2$]/[O$_2$,\,no\,WGH]} \\
>27 & 1.37 & 0.68 & 10.16 & 1.10 \\
>30 & 2.27 & 0.83 & 8.19 & 1.78 \\
>40 & 3.30 & 1.46 & 1.89 & 5.15 \\
>50 & 2.48 & 1.25 & 1.85 & 4.16 \\
>60 & 1.76 & 1.08 & 1.55 & 3.11 \\
>70 & 1.32 & 1.18 & 1.25 & 3.11 \\
>80 & 1.10 & 1.32 & 1.07 & 3.00 \\
>90 & 1.01 & 1.00 & 0.98 & 0.99 \\
\hline
	\end{tabular}
\end{table}

\begin{table}
	\centering
\caption{Calculated abundance ratio for the CO$_2$ and H$_2$O molecules in the icy mantles: results for simulations with WGH (eight $T_{\rm CR}$ thresholds) versus results without WGH. The latter are provided within the table.}
\label{tab-h2o}
	\begin{tabular}{lcccc}
  \hline
 & \multicolumn{2}{c}{CO$_2$ ice} & \multicolumn{2}{c}{H$_2$O ice} \\
Abundance, & 1.5 Myr & 2.5 Myr & 1.5 Myr & 2.5 Myr \\
no WGH, cm$^{-3}$ & 4.65E-05 & 8.23E-05 & 1.08E-04 & 7.62E-05 \\
\hline
($T_{\rm CR}$, K) & \multicolumn{2}{c}{[CO$_2$]/[CO$_2$,\,no\,WGH]} & \multicolumn{2}{c}{[H$_2$O]/[H$_2$O,\,no\,WGH]} \\
>27 & 1.02 & 1.00 & 1.01 & 1.08 \\
>30 & 0.52 & 0.70 & 1.32 & 1.51 \\
>40 & 0.27 & 0.54 & 1.47 & 1.86 \\
>50 & 0.44 & 0.66 & 1.38 & 1.61 \\
>60 & 0.64 & 0.78 & 1.25 & 1.38 \\
>70 & 0.82 & 0.91 & 1.13 & 1.18 \\
>80 & 0.92 & 0.97 & 1.05 & 1.06 \\
>90 & 1.01 & 0.99 & 1.00 & 1.01 \\
\hline
	\end{tabular}
\end{table}

\begin{table}
	\centering
\caption{Calculated abundance ratio for the N atom and the N$_2$ molecule in gas phase: results for simulations with WGH (eight $T_{\rm CR}$ thresholds) versus results without WGH. The latter are provided within the table.}
\label{tab-n}
	\begin{tabular}{lcccc}
  \hline
 & \multicolumn{2}{c}{N gas} & \multicolumn{2}{c}{N$_2$ gas} \\
Abundance, & 1.5 Myr & 2.5 Myr & 1.5 Myr & 2.5 Myr \\
no WGH, cm$^{-3}$ & 1.60E-07 & 1.42E-09 & 6.82E-07 & 1.97E-09 \\
\hline 
($T_{\rm CR}$, K) & \multicolumn{2}{c}{[N]/[N,\,no\,WGH]} & \multicolumn{2}{c}{[N$_2$]/[N$_2$,\,no\,WGH]} \\
>27 & 3.16 & 3.42 & 17.03 & 2.92 \\
>30 & 4.29 & 7.96 & 29.56 & 8.65 \\
>40 & 3.53 & 23.69 & 24.21 & 33.53 \\
>50 & 3.02 & 16.66 & 12.29 & 22.08 \\
>60 & 2.16 & 6.28 & 5.10 & 7.50 \\
>70 & 1.50 & 2.89 & 2.28 & 3.18 \\
>80 & 1.15 & 2.07 & 1.33 & 2.19 \\
>90 & 1.02 & 1.01 & 1.04 & 1.02 \\
\hline
	\end{tabular}
\end{table}

\begin{table}
	\centering
\caption{Calculated abundance ratio for the NH$_3$ and N$_2$ molecules in the icy mantles: results for simulations with WGH (eight $T_{\rm CR}$ thresholds) versus results without WGH. The latter are provided within the table.}
\label{tab-nh3}
	\begin{tabular}{lcccc}
  \hline
 & \multicolumn{2}{c}{NH$_3$ ice} & \multicolumn{2}{c}{N$_2$ ice} \\
Abundance, & 1.5 Myr & 2.5 Myr & 1.5 Myr & 2.5 Myr \\
no WGH, cm$^{-3}$ & 1.51E-05 & 1.37E-05 & 2.89E-05 & 3.04E-05 \\
\hline 
($T_{\rm CR}$, K) & \multicolumn{2}{c}{[NH$_3$]/[NH$_3$,\,no\,WGH]} & \multicolumn{2}{c}{[N$_2$]/[N$_2$,\,no\,WGH]} \\
>27 & 1.21 & 2.10 & 0.56 & 0.71 \\
>30 & 1.19 & 3.26 & 0.27 & 0.44 \\
>40 & 1.15 & 2.50 & 0.41 & 0.59 \\
>50 & 1.15 & 1.76 & 0.69 & 0.80 \\
>60 & 1.12 & 1.33 & 0.87 & 0.91 \\
>70 & 1.06 & 1.13 & 0.95 & 0.96 \\
>80 & 1.02 & 1.06 & 0.99 & 0.98 \\
>90 & 1.03 & 1.02 & 0.99 & 1.00 \\
\hline
	\end{tabular}
\end{table}

\bsp	
\label{lastpage}
\end{document}